\documentclass[11pt]{amsart}
\usepackage{geometry}                
\usepackage{amsmath}
\usepackage{amsthm}
\usepackage{graphicx}
\usepackage{amssymb}
\usepackage[title]{appendix}
\usepackage{extarrows}
\usepackage{ulem}
\usepackage{xcolor}

\title{Limiting free energy per particle for Ising Model by approximating its functional integral}
\author{Rong Qiang Wei}
\address{College of Earth and Planetary Sciences, University of Chinese Academy of Sciences, Beijing, PRC, 100049}
\email{wrq1973@ucas.edu.cn}
\date{}

\begin{document}
\maketitle

\begin{abstract}
    There have been a lot of methods aimed at studying the limiting free energy per particle (LFEPP) for 3-dimensional (3D) Ising model in absence of an external magnetic field. These methods are elegant, but most of them are complicated and often require specialized knowledge and special skills. Here we approximate the LFEPP for Ising model from its functional integral using classic mathematical-physical methods. The resulting LFEPPs for 1-dimensional (1D) to 3D Ising model have similar structures and forms. We then verify that these LFEPPs are correct for two limiting cases of the 1D and 2-dimensional (2D) models, as well as for the critical inverse temperature $z_c$ of the 2D model. Based on these verifications, we derive naturally the LFEPP and the $z_c$ ($\approx 0.21\sim 0.22$) for the 3D model. Furthermore, we suggest similar LFEPPs for 1D-3D Ising models with an external magnetic field, although they are too complicated.

\end{abstract}

{\hspace{2.2em}\small Keywords:}

{\hspace{2.2em}\tiny Ising model, Limiting free energy, Functional integral, Critical inverse temperature}

\setcounter{page}{0}
{\tiny \tableofcontents} 

\section{Introduction}\label{intro}

The famous Ising model looks very simple, which consists of a lattice with a binary magnetic polarity (or "spin") assigned to each point. The nearest-neighbor Ising model without an external magnetic field in $D$-dimensions ($D=1,2,3,...$) is defined in terms of the following Hamiltonian (e.g., Huang, 1987),

\begin{equation}\label{eq1}
\mathcal{H} =  -\frac{1}{2}\sum\limits_{i,j = 1}^N {{K_{ij}}} {s_i}{s_j}
\end{equation}
where, $i$ and $j$ are the sites $\bf{r}_i$ and $\bf{r}_j$ of a $D$-dimensional hyper cubic lattice with $N$ sites, respectively. $s_i=\pm 1$ are the two possible states of the $z$-components of spins localized at the lattice sites. $K_{ij}$ denotes the exchange interaction between spins localized at $\bf{r}_i$ and $\bf{r}_j$, 

\begin{equation}\label{eq2}
{K_{ij}} = \left\{ {\begin{array}{ll}
z&{{\rm{if \hspace{0.5em}}}i{\rm{\hspace{0.5em}and\hspace{0.5em}}}j{\rm{\hspace{0.5em} are\hspace{0.2em} the \hspace{0.2em}nearest\hspace{0.2em} neighbors}}}\\
0&{{\rm{otherwise}}}
\end{array}} \right.
\end{equation}
where $z=\beta\epsilon=\frac{\epsilon}{kT}$, $\epsilon$ the interaction energy, $T$ the temperature, and $k$ Boltzmann constant.

Such a simple model has played an important role in the theory of ferromagnetism, phase transitions and many other physical branches (Fang et al., 2022), and it has been found new applications in many areas of science, e.g., geophysics, neuroscience, and even voter models (Jimenez et al., 2007; Viswanathan et al., 2022). 

Most thermodynamic functions of this model depend on the evaluation of its limiting free energy per particle (LFEPP). For 1-dimensional (1D) Ising model with periodic boundary condition, the exact LFEPP, $-\frac{\psi}{kT}$, is,

\begin{equation}\label{KW1}
	-\frac{\psi}{kT}=
		\frac{1}{2}\frac{1}{2\pi}\int_0^{2\pi}{\log \left( \cosh 2z-\sinh 2z\cos \omega _1 \right)}\mathrm{d}\omega _1
\end{equation}
where we use the LFEPP in terms of integral from Berlin and Kac (1952) rather than that from Kramers and Wannier (1941) for comparison later. So does the following 2-dimensional (2D) Ising model.  

For 2D Ising model imposed on periodic boundary condition, the famous LFEPP which was evaluated by Onsager (1944) is, 

\begin{equation}\label{Onsager1}
	-\frac{\psi}{kT}=
	\frac{1}{2}\frac{1}{(2\pi)^2}\iint\limits_0^{2\pi}{\log \left[\cosh^2 2z-\sinh 2z(\cos \omega _1+\cos \omega _2) \right]}\mathrm{d}\omega _1\mathrm{d}\omega _2
\end{equation}

From (\ref{Onsager1}), the critical inverse temperature $z_c$ for 2D model can be obtained easily by solving the equation that $\cosh^2 2z-2\sinh 2z=0$.

However,  there is still no an accepted LFEPP for 3-dimensional (3D) Ising model, although for this purpose there have been lots of methods or approaches which are valid for 1D, 2D and/or high-dimensional ($D>3$) models. The most famous ones are Mean-Field Theory (e.g., Bragg and William, 1934, 1935; Williams, 1935; Landau, 1937), Transfer Matrix Method (Onsager, 1944), $\varphi^4-$ Theory (Ginzburg and Landau, 1950), Variational Calculation (Thompson, 1965), Conformal bootstrap (El-Showk et al., 2012; 2014). Especially the Transfer Matrix Method and the similar ones are popular in the later studies (e.g., Zhang, 2007), although their results are questionable (e.g., Wu et al., 2008; Fisher and Perk, 2016; Perk, 2013). Transfer Matrix Method requires specialized and abstract knowledge of spinor algebra or operator algebra, which is unfamiliar to most non-professionals. In 2015, Kocharovsky and Kocharovsky (2015) presented a method of the recurrence equations for partial contractions for 3D model, and they said that "Towards an exact solution for the three-dimensional Ising model". However, it can also be found that their method is too complicated and specialized.   

 Here we study the LFEPP of Ising Model from its corresponding functional integral with classic mathematical-physical methods. Some interesting results are obtained without usnig specialized knowledge and/or special skills, as shown in section \ref{HS_0h}-\ref{approx_general} and the relevant appendices. In section 2 we introduce the functional integral of the Ising model and propose the general LFEPPs of 1D–3D models. Then these LFEPPs in three special cases will be studied in section 3. In section 4 we further approximate the LFEPPs in the 2D-3D Ising model, estimate their critical inverse temperatures, and propose the LFEPPs for 1D–3D Ising models with an external magnetic field. The main conclusions are summarized and discussed in the final section.

\section{Functional integral and the LFEPP of Ising model}\label{HS_0h}

The partition function for the Ising model (\ref{eq1}) is (e.g., Ginzburg and Landau, 1950),

\begin{equation}\label{eq3}
\mathcal{Z} =\sum_{\left\{ s_i=\pm 1 \right\}}{\exp \left( \frac{1}{2}\sum_{ij}{K_{ij}s_is_j} \right)}
\end{equation}

Applying the Hubbard-Stratonovich transformation (Ginzburg and Landau, 1950; Amit et al., 2005; Kopietz et al., 2010; Also in Appendix \ref{ApendH}), the partition function can be transformed into,

\begin{align}\label{eq4}
	\begin{aligned}
		\mathcal{Z}& =\sum_{\left\{ s_i=\pm 1 \right\}}\left[ \frac{\det  \mathbf{K}}{\left( 2\pi \right) ^N} \right] ^{1/2}\int_{-\infty}^{+\infty}\int_{-\infty}^{+\infty}{\cdots}\int_{-\infty}^{+\infty}\prod_{k=1}^N\mathrm{d}\phi _k\\
		&\hspace{7em}\times\exp \left( -\frac{1}{2}\sum_{ij}{\phi _iK_{ij}\phi _j+\sum_{ij}{s_iK_{ij}\phi _j}} \right)\\
	\end{aligned}
\end{align}
where $K_{ij}$ is a circulant matrix when an appropriate periodicity is used on the boundary of Ising model, and it is assumed to be positive defined when $N\rightarrow \infty$. Details including eigenvalues and eigenvectors of $K_{ij}$ can be found in Appendix \ref{ApendA} ((\ref{eig_1D})-(\ref{eig_3D}),(\ref{eig_vector})). 

(\ref{eq4}) expresses the partition function of the Ising model in terms of an $N$-dimensional integral over variables $\phi_i$. The physical meaning of the $\phi_i$ is the fluctuating magnetization per site, for its expectation value is simply the
magnetization per site. The infinite-dimensional ($N\rightarrow\infty$) integral from (\ref{eq4})  is called the functional integral of the Ising model, which is an exact transcription of the original model (\ref{eq3}) (e.g., Ginzburg and Landau, 1950; Amit et al., 2005; Kopietz et al., 2010). 

In the second row of (\ref{eq4}), it can be found that only $\sum_{ij}{s_iK_{ij}\phi _j}$ in the exponential has a dependence on the $s_i$. Therefore we can calculate that $L=\sum_{\left\{ s_i=\pm 1 \right\}} \exp \left( \sum_{ij}{s_iK_{ij}\phi _j} \right)$ as follows,  

\begin{align}\label{eq5}
	\begin{aligned}
		L&=\sum_{\left\{ s_i=\pm 1 \right\}}{\exp \left( \sum_p{s_iV_{ip}V_{pi}K_{ij}V_{jp}V_{pj}\phi _j} \right)}
		\\
		&=\sum_{\left\{ s_i=\pm 1 \right\}}{\prod_p{\prod_i{\exp \left[ s_i\left( V_{ip}\lambda _py_p \right) \right]}}}
		\\
		&=\prod_p\prod_i{\left[\frac{ \exp \left( V_{ip}\lambda _py_p \right) +\exp \left( -V_{ip}\lambda _py_p \right)}{2}\right]}
		\\
		&=\prod_p{\exp \left\{ \sum_i{\log \left[ \cosh \left( V_{ip}\lambda _py_p \right) \right]} \right\}}
		\end{aligned}
\end{align}
where $V_{ij}$ are the real, orthogonal, unity eigenvectors corresponding to eigenvalues $\lambda_p$ of $K_{ij}$ ((\ref{eig_vector})); "2" can be understood as a normalized factor (to unity). Einstein summation convention is used in (\ref{eq5}), for example, $y_p=V_{pj}\phi _j=\sum_j{V_{pj}\phi _j}$.

Inserting (\ref{eq5}) into (\ref{eq4}) and noticing the Jacobian unity, now the partition function (\ref{eq4}) is,

\begin{align}\label{eq6}
	\begin{aligned}
	\mathcal{Z} &= \left[ \frac{\det\mathbf{K}  }{\left( 2\pi \right) ^N} \right] ^{1/2}\int_{-\infty}^{+\infty}\int_{-\infty}^{+\infty}\dotsi\int_{-\infty}^{+\infty}\prod_{p=1}^N \exp \left( -\frac{1}{2}\lambda _py_{p}^{2} \right)\\
	&\hspace{13.5em}\times  \exp \left\{ \sum_i{\log \left[ \cosh \left( V_{ip}\lambda _py_p \right) \right]} \right\} \mathrm{d}y_p\\
    &= \left[ \frac{\det\mathbf{K}}{\left( 2\pi \right) ^N} \right] ^{1/2}\prod_{p=1}{\int_{-\infty}^{+\infty}{\exp \left( -\frac{1}{2}\lambda _py_{p}^{2} \right) \times \exp \left\{ \sum_i{\log \left[\cosh \left( V_{ip}\lambda _py_p \right) \right]} \right\} }}\mathrm{d}y_p\\
		&= \left[ \frac{\det\mathbf{K}}{\left( 2\pi \right) ^N} \right] ^{1/2}\prod_{p=1} I_p
	\end{aligned}
\end{align}

Expression (\ref{eq6}) is equivalent to (\ref{eq4}), and we call it also the functional integral of the Ising model. Once $I_p$ is calculated, the LFEPP, $-\frac{\psi}{kT}$, is,

\begin{equation}\label{eq8}
	-\frac{\psi}{kT}=\lim_{N\rightarrow \infty}\frac{\log\mathcal{Z}}{N} 
\end{equation}

According to Appendix \ref{ApendB}, $I_p$ in (\ref{eq6}) can be transformed into two integrals of (\ref{I_p_integral}) in the following,

\begin{align}\label{I_p_integral}
	\begin{aligned}
		I_p=
		\left\{ \begin{array}{ll}
			N^{1/2}\int_{-\infty}^{+\infty}{\exp}\left\{ N\left[ -\frac{1}{2}\lambda _1t^2+\log \left( \cosh \left( \lambda _1t \right) \right) \right] \right\} \mathrm{d}t&		p=1\\
			N^{1/2}\int_{-\infty}^{+\infty}{\exp}\left\{ N\left[ -\frac{1}{2}\lambda _pt^2+\frac{2}{\pi}\int_0^{\pi /2}{\log \left[ \cosh \left( \sqrt{2}\lambda _pt\sin \xi \right) \right] \mathrm{d}\xi} \right] \right\} \mathrm{d}t&		p>1\\
		\end{array} \right. 
\end{aligned}
\end{align}

These integrals have been already calculated with Laplace's asymptotic method in the Appendix \ref{ApendB}. For that $p=1$, 

\begin{align}\label{eq18_zw}
	\begin{aligned}
		I_1	\sim \left\{
		\begin{array}{ll}
			2\exp \left\{ N\left[ -\frac{1}{2}\lambda _1 t_{s_1}^{2}+\log \left( \cosh \left( \lambda _1 t_{s_1} \right) \right) \right] \right\} \sqrt{\frac{2\pi}{\vert  -\lambda _1+\lambda _{1}^{2}{\rm sech} ^2\left( \lambda _1 t_{s_1} \right) \vert }}& t_{s1}\neq 0\\
			2 \sqrt{\frac{\pi}{2\vert  -\lambda _1+\lambda _{1}^{2} \vert }}& t_{s1}=0
		\end{array}
		\right.		
	\end{aligned}
\end{align}

And for that $p>1$, 

\begin{align}\label{eq14_zw}
	\begin{aligned}
		I_p	\sim \left\{
		\begin{array}{ll}
			2 \exp \left[ Nf(\lambda_p,t_{sp})\right] \sqrt{\frac{2\pi}{\left| -\lambda _p+\mathcal{I} ^{''}\left( \lambda _p,t_{sp} \right) \right|}}& t_{sp}\neq 0\\
			2 \exp \left[ Nf(\lambda_p,0)\right] \sqrt{\frac{\pi}{2 \left| -\lambda _p+\mathcal{I} ^{''}\left( \lambda _p,0 \right) \right|}}& t_{sp}=0
		\end{array}
		\right.			
	\end{aligned}
\end{align}
where $	f\left( \lambda _p,t \right) =-\frac{1}{2}\lambda _pt^2+\frac{2}{\pi} \int_0^{\pi /2}{\log \left[ \cosh \left( \sqrt{2}\lambda _pt\sin \xi \right) \right] \mathrm{d}\xi}=-\frac{1}{2}\lambda _pt^2+\mathcal{I}(\lambda_p, t)$, and details of $f(\lambda_p,t)$, $\mathcal{I} ^{''}\left( \lambda _p,t \right)$, $t_{s1}$ and $t_{sp}$ are in Appendix \ref{ApendB}. 

Assuming that $t_{sp}\neq 0$ and inserting $I_p$ above into (\ref{eq6}), we can solve the functional integral of the Ising model, or its equivalence (\ref{eq4}). Finally, we obtain the partition function of $N$-sites Ising model in the following,

\begin{align}\label{eq19}
	\begin{aligned}
		\mathcal{Z} & =\left[ \frac{\det \mathbf{K}}{\left( 2\pi \right) ^N} \right] ^{1/2}  I_1\prod_{p=2}{ I_p } \\
		&\sim  2\exp \left\{ N\left[ -\frac{1}{2}\lambda _1t_{s1}^{2}+\log \left( \cosh \left( \lambda _1t_{s1} \right) \right) \right] \right\} \left| 1-\lambda _1{\rm sech}^2\left( \lambda _1t_{s1} \right)  \right| ^{-1/2}\\
		&\hspace{4em} \times \prod_{p=2}{ 2\exp[Nf(\lambda_p,t_{sp})] \left| -1+\frac{\mathcal{I} ^{''}\left( \lambda _p,t_{sp} \right)}{\lambda _p} \right| ^{-1/2}}
	\end{aligned}
\end{align}
where $\det \mathbf{K} =\prod_{p=1} \lambda _p$.
      
And, 
      
\begin{align}\label{eq19_2023}
	\begin{aligned}
     \frac{\log \mathcal{Z}}{N}&=\frac{\log 2}{N}+\left\{ -\frac{1}{2}\lambda _1t_{s1}^{2}+\log \left[\cosh \left(t_{s1}\lambda_1\right) \right] \right\} -\frac{1}{2N}\log \left| 1-\lambda _1{\rm sech}^2\left( \lambda _1t_{s1} \right) \right| \\
     &\hspace{2em} +\frac{(N-1)\log 2}{N}+\sum_{p=2}^N {f(\lambda_p,t_{sp})} +\frac{1}{N}\sum_{p=2}^{N}{\left[-\frac{1}{2}\log \left| -1+\frac{\mathcal{I} ^{''}\left( 2\frac{\lambda _p}{2},t_{sp} \right)}{2\frac{\lambda _p}{2}}\right|\right]}
   \end{aligned}
\end{align}

Now we can get easily the LFEPP from (\ref{eq19_2023}) when $N\rightarrow \infty$ according to Appendix \ref{ApendC}. It should be pointed out that expression (\ref{eq19}) and (\ref{eq19_2023}) are applicable to any dimensional Ising models.  However, the following are only applicable to 1-3D models, because the proofs in Appendix \ref{ApendC} are only for them. According to (\ref{Sf_1D}), (\ref{Sf_2D}) and (\ref{Sf_3D}), the LFEPPs of 1-3D Ising model can be expressed as follows.

For 1D Ising model, (\ref{eq19_2023}) reads,

\begin{align}\label{1DFE}
	\begin{aligned}
	-\frac{\psi}{kT}=\lim_{N\rightarrow \infty} \frac{\log \mathcal{Z}}{N}&= -zt_{s1}^{2}+\log \left[\cosh \left( 2zt_{s1} \right) \right] + \log 2 +\sum_{p=2}^\infty {f(\lambda_p,t_{sp})} \\
	&\hspace{1em}  -\frac{1}{2}\frac{1}{2\pi}\int_0^{2\pi}{\log \left| -1+\frac{\mathcal{I} ^{''}\left( 2z\cos \omega _1,t_{sp} \right)}{2z\cos \omega _1} \right| \mathrm{d}\omega _1}
	\end{aligned}
\end{align}

For 2D Ising model, (\ref{eq19_2023}) reads,

\begin{align}\label{2DFE}
	\begin{aligned}
		-\frac{\psi}{kT}&=-2zt_{s1}^{2}+\log \left[\cosh \left( 4zt_{s1} \right)\right] + \log 2 +\sum_{p=2}^\infty {f(\lambda_p,t_{sp})}
		\\
		&\hspace{1em} -\frac{1}{2}\frac{1}{\left( 2\pi \right) ^2}\iint\limits_0^{2\pi}{\log \left| -1+\frac{\mathcal{I} ^{''}\left[ 2z\left( \cos \omega _1+\cos \omega _2 \right),t_{sp} \right]}{2z\left( \cos \omega _1+\cos \omega _2 \right)} \right| \mathrm{d}\omega _1\mathrm{d}\omega _2}
	\end{aligned}
\end{align}

For 3D Ising model, (\ref{eq19_2023}) reads,

\begin{align}\label{3DFE}
	\begin{aligned}
		-\frac{\psi}{kT}&=-3zt_{s1}^{2}+\log \left[\cosh \left( 6zt_{s1} \right) \right] +\log 2 + \sum_{p=2}^\infty {f(\lambda_p,t_{sp})}\\
		&\hspace{1em} -\frac{1}{2}\frac{1}{\left( 2\pi \right) ^3}\iiint\limits_0^{2\pi}{\log \left| -1+ \frac{\mathcal{I} ^{''}\left[ 2z\left( \cos \omega _1+\cos \omega _2+\cos \omega _3 \right) ,t_{sp}\right]}{2z\left( \cos \omega _1+\cos \omega _2+\cos \omega _3 \right)} \right|\,\,\mathrm{d}\omega _1\mathrm{d}\omega _2\mathrm{d}\omega _3}
	\end{aligned}
\end{align}

It can be found that the integral parts in (\ref{1DFE}), (\ref{2DFE}) and (\ref{3DFE}) are in terms of the extended Joyce and Zucker's logarithmic integral (Joyce and Zucker, 2001). From the logarithmic integral in (\ref{2DFE}) and (\ref{3DFE}), $z_c$ for 2D and 3D Ising model can be obtained by solving the equation that $\mathcal{I} ^{''}\left( 4z,t_{sp} \right)=4z$ and that $\mathcal{I} ^{''}\left( 6z,t_{sp} \right)=6z$, respectively.  

\section{LFEPP of Ising model in some special cases}\label{SpecialCase}

To our limited knowledge, we infer that $\mathcal{I}(\lambda_p, t)$ ($=\frac{2}{\pi} \int_0^{\pi /2}{\log \left( \cosh \left( \sqrt{2}\lambda_pt\sin \xi \right) \right) \mathrm{d}\xi}$) is a special function, resulting in that we can not obtain easily $t_{sp}$ and further evaluating of $f(\lambda_p,t_{sp})$ and  $\mathcal{I}^{''}(\lambda_p, t_{sp})$. In the following some approximate LFEPPs to (\ref{eq19_2023})-(\ref{3DFE}) are taken into account. Since we only consider the behaviour of $t$ near $t_{sp}$, $t$ is finite.  

\subsection{$z \rightarrow 0$}\label{SpecialCase1}

In this case, $\sqrt{2}\lambda _pt\ (=\sqrt{2} z \lambda_p^\prime t)$ 
is in some sense small and since $$
\log \cosh \left( x \right) \approx \frac{x^2}{2}-\frac{x^4}{12}\hspace{3em} (x\rightarrow 0)
$$

\begin{align}\label{eq21}
	\begin{aligned}	
		f\left( \lambda _p,t \right) &=-\frac{1}{2}\lambda _pt^2+\mathcal{I}(\lambda_p,t)\\
		&\approx -\frac{1}{2}\lambda _pt^2+\frac{\left( \lambda _pt \right) ^2}{2}-\frac{\left( \lambda _pt \right) ^4}{8}
	\end{aligned}
\end{align}

and,

\begin{align}\label{eq22}
	\begin{aligned}
		\begin{cases}
			\frac{{\rm d}f}{{\rm d}t}=-\lambda _pt+\lambda _{p}^{2}t-\frac{\lambda _{p}^{4}}{2}t^3\\
			\frac{{\rm d}^2f}{{\rm d}t^2}=-\lambda _p+\lambda _{p}^{2}-\frac{3\lambda _{p}^{4}}{2}t^2\\
		\end{cases}
	\end{aligned}
\end{align}

From Eq. (\ref{eq22}) we know that $t_{sp}=0$. Since $t_{sp}=0$, 

\begin{align}\label{eq23}
	\begin{aligned}
		\begin{cases}
			f\left( \lambda _p,t_{sp} \right)=0\\
			\frac{{\rm d}^2f}{{\rm d}t^2}|_{t=t_{sp}}=-\lambda _p+\mathcal{I}^{''}(\lambda_p, t_{sp})=-\lambda _p+\lambda _{p}^{2}\ (<0)\\
		\end{cases}
	\end{aligned}
\end{align}

With Laplace's method and directly from Eq. (\ref{eq19_2023}), we obtain (Note that $t_{s1}=0$),

\begin{equation}\label{eq24}
	\frac{\log \mathcal{Z}}{N}=\log 2+ \frac{1}{N}\sum_{p=2}^{N}{\left(-\frac{1}{2}\log \left| -1+2\frac{\lambda _p}{2}\right|\right)}
\end{equation}

Finally, we obtain the LFEPP of Ising model according to Appendix C. Namely,

For 1D Ising model,

\begin{equation}\label{eq25}
	-\frac{\psi}{kT}=\lim_{N\rightarrow \infty}\frac{\log\mathcal{Z}(\lambda_p/2)}{N}= \log 2-\frac{1}{2}\frac{1}{2\pi}\int_0^{2\pi}{\log \left( 1-2z \cos \omega _1  \right)  \mathrm{d}\omega _1}
\end{equation}

For 2D Ising model,

\begin{equation}\label{eq26}
	-\frac{\psi}{kT}=\log 2-\frac{1}{2}\frac{1}{\left( 2\pi \right) ^2}\iint\limits_0^{2\pi}{\log  \left[1-2z\left( \cos \omega _1+\cos \omega _2 \right)  \right]   \mathrm{d}\omega _1}\mathrm{d}\omega _2
\end{equation}

Using that $\cosh2z\rightarrow 1$ and that $\sinh2z\rightarrow 2z$ when $z\rightarrow 0$, we can transform (\ref{KW1}) and (\ref{Onsager1}) into the following,

\begin{equation}\label{KW1_1}
	-\frac{\psi}{kT}=
	\frac{1}{2}\frac{1}{2\pi}\int_0^{2\pi}{\log \left( 1- 2z\cos \omega _1 \right)}\mathrm{d}\omega _1
\end{equation}

\begin{equation}\label{Onsager1_1}
	-\frac{\psi}{kT}=
	\frac{1}{2}\frac{1}{(2\pi)^2}\iint\limits_0^{2\pi}{\log \left[1 - 2z(\cos \omega _1+\cos \omega _2) \right]}\mathrm{d}\omega _1\mathrm{d}\omega _2
\end{equation}

It can be found that (\ref{KW1_1}) and (\ref{eq25}) are nearly the same, except the minus sign before the integral representation and "$\log 2$" which will be discussed later. So do (\ref{Onsager1_1}) and (\ref{eq26}). Therefore, we can infer the LFEPP for 3D Ising model from (\ref{eq24}) when $z\rightarrow 0$ in the following,

\begin{align}\label{3D_small_z}
	\begin{aligned}	
		-\frac{\psi}{kT} = \log 2-\frac{1}{2}\frac{1}{\left( 2\pi \right) ^3}\iiint\limits_0^{2\pi}{\log \left[ 1-2z\left( \cos \omega _1+\cos \omega _2+\cos \omega _3 \right) \right]\,\,\mathrm{d}\omega _1\mathrm{d}\omega _2\mathrm{d}\omega _3}
	\end{aligned}
\end{align}

\subsection{$z \rightarrow \infty$}\label{SpecialCase2}

Let us assume that $z$ is in some sense large. Although in this case that $\lambda_p t=z \lambda_p^{\prime} t\rightarrow\infty$, $\frac{\log[\cosh(\lambda_p t)]}{N}$ in (\ref{eq12}) can also be omitted according to Appendix \ref{ApendE}. 

When $z\rightarrow \infty$, $t_{s1}\rightarrow 1$; $\mathcal{I}(\lambda_p, t)\rightarrow \frac{2\sqrt{2}\lambda_p}{\pi}t-\log 2$. Thus,

\begin{equation}\label{eq28n}
	\left\{
	\begin{array}{ll}	
		f\left( \lambda _p,t \right) &\sim -\frac{1}{2}\lambda _pt^2+\frac{2\sqrt{2}\lambda_p}{\pi}t-\log 2\\
		\frac{{\rm d}f}{{\rm d}t} &\sim -\lambda _pt+\frac{2\sqrt{2}\lambda_p}{\pi}\\
		\frac{{\rm d}^2f}{{\rm d}t^2} &\sim -\lambda _p
	\end{array}
	\right.
\end{equation}

From (\ref{eq28n}) we know that $t_{sp}\rightarrow \frac{2\sqrt{2}}{\pi}$, $\mathcal{I}^{''}(\lambda_p, t_{sp})\rightarrow 0$ and that $f\left( \lambda _p,t_{sp} \right) \sim \frac{4}{\pi^2}\lambda_p-\log 2$. Therefore,

\begin{align}\label{eq29n}
	\begin{aligned}
		\frac{\log \mathcal{Z}}{N}&=\frac{\log 2}{N}+\left\{ -\frac{1}{2}\lambda _1+\log \left[\cosh \left(\lambda_1\right) \right] \right\} -\frac{1}{2N}\log \left| 1-\lambda _1{\rm sech}^2\left( \lambda _1\right) \right| \\
		&\hspace{2em} +\frac{(N-1)\log 2}{N}+\sum_{p=2}^N {\left(\frac{4}{\pi^2}\lambda_p-\log 2\right)} 
	\end{aligned}
\end{align}

Since $\lambda_p$ ($=z\lambda_p^\prime$) is large, 

\begin{align}\label{eq30n}
	\begin{aligned}
		\frac{\log \mathcal{Z}}{N}&\rightarrow \frac{\lambda _1}{2}+\frac{4}{\pi^2}z\sum_{p=2}^N  {\lambda_p^\prime} 
	\end{aligned}
\end{align}

From (\ref{eq30n}) and Appendix \ref{ApendD}, we can get,

\begin{equation}\label{Asymptotic_1D}
-\frac{\psi}{kT}= \left( 1-\frac{8}{\pi ^2} \right) z
\end{equation}

for 1D Ising model. 

For 2D Ising model,

\begin{equation}\label{Asymptotic_2D}
	-\frac{\psi}{kT}= 2\left( 1-\frac{8}{\pi ^2} \right) z
\end{equation}

On the other hand, for 1D Ising model, according to (\ref{KW1}) \footnote{Note that $\int_0^{2\pi}{\log \left( 1-\cos \omega _1 \right)}\mathrm{d}\omega _1=-2\pi \log 2$.}, 

\begin{align}\label{pre_Asymptotic_1D}
	\begin{aligned}	
		-\frac{\psi}{kT}&=\frac{1}{2}\frac{1}{2\pi}\int_0^{2\pi}{\log \left( \cosh 2z-\sinh 2z\cos \omega _1 \right)}\mathrm{d}\omega _1
		\\
		&\,\,       =\frac{1}{2}\frac{1}{2\pi}\int_0^{2\pi}{\log \left( \frac{e^{2z}}{2}-\frac{e^{2z}}{2}\cos \omega _1 \right)}\mathrm{d}\omega _1 
		\\
		&\,\,      =z\left( 1-\frac{\log 2}{z} \right)\\
		& \,\,     \rightarrow z \hspace{2em} (z\rightarrow \infty)
	\end{aligned}
\end{align}

According to (\ref{Onsager1}) for 2D Ising model,

\begin{align}\label{pre_Asymptotic_2D}
	\begin{aligned}	
		-\frac{\psi}{kT}&=\frac{1}{2}\frac{1}{(2\pi )^2}\int_0^{2\pi}{\int_0^{2\pi}{\log \left[ \cosh ^22z-\sinh 2z(\cos \omega _1+\cos \omega _2) \right]}}\mathrm{d}\omega _1\mathrm{d}\omega _2
		\\
		&\,\,      =\frac{1}{2}\frac{1}{(2\pi )^2}\int_0^{2\pi}{\int_0^{2\pi}{\log \left[ \frac{e^{4z}}{4}-\frac{e^{2z}}{2}(\cos \omega _1+\cos \omega _2) \right]}}\mathrm{d}\omega _1\mathrm{d}\omega _2
		\\
		&\,\,      =2z\left( 1-\frac{\log 2}{2z} \right) \\
		& \rightarrow 2z (z\rightarrow \infty)
	\end{aligned}
\end{align}

It can be found that (\ref{Asymptotic_1D}) and  (\ref{pre_Asymptotic_1D}) are of the same order as $z\rightarrow \infty$; So do (\ref{Asymptotic_2D}) and (\ref{pre_Asymptotic_2D}).

Clearly, for 3D Ising model when $z$ is large, we have,

\begin{equation}\label{Asymptotic_3D}
	-\frac{\psi}{kT} =3\left( 1-\frac{8}{\pi ^2} \right) z    
\end{equation}

\subsection{$\sqrt{2} z\lambda_p^\prime t \leq \pi$}\label{SpecialCase3}

From the previous sections, we have,

\begin{align}\label{specialcase3_1}
	\left\{
	\begin{aligned}	
		f\left( \lambda _p,t \right) &=-\frac{1}{2}\lambda _pt^2+\frac{2}{\pi}\int_0^{\pi /2}{\log \left( \cosh \left( \sqrt{2}\lambda _pt\sin \xi \right) \right) \mathrm{d}\xi}
		\\
		\frac{{\rm d}f}{{\rm d}t} &= -\lambda _pt+\frac{2}{\pi}\int_0^{\pi /2} (\sqrt{2}\lambda _p \sin \xi) {\ \tanh \left( \sqrt{2}\lambda _pt\sin \xi \right) \mathrm{d}\xi}\\
		\frac{{\rm d}^2f}{{\rm d}t^2} &= -\lambda _p+\frac{2}{\pi}\int_0^{\pi /2} (\sqrt{2}\lambda _p \sin \xi)^2 {\ {\rm sech}^2 \left( \sqrt{2}\lambda _pt\sin \xi \right) \mathrm{d}\xi}
	\end{aligned}
	\right.
\end{align}

From that ${\rm d}f/{\rm d}t=0$, we have,

\begin{align}\label{specialcase3_2}
	\begin{aligned}	
		 \lambda _pt^2=\frac{2}{\pi}\int_0^{\pi /2} (\sqrt{2}\lambda _p t \sin \xi) {\ \tanh \left( \sqrt{2}\lambda _pt\sin \xi \right) \mathrm{d}\xi}
	\end{aligned}
\end{align}

Let $x=\sqrt{2}\lambda _p t$, and (\ref{specialcase3_2}) means,

\begin{align}\label{specialcase3_3}
	\begin{aligned}	
		\frac{x^2}{2\lambda_p}=f(x)&=\frac{2}{\pi}\int_0^{\pi /2} (x\sin \xi) {\ \tanh \left(x\sin \xi \right) \mathrm{d}\xi}\\
		&=\frac{2}{\pi}\int_0^{\pi /2} G(x\sin \xi)\mathrm{d}\xi
	\end{aligned}
\end{align}

(\ref{specialcase3_3}) is the Schl$\ddot{\mathrm{o}}$milch integral equation, and it has one solution when $-\pi\leq x \leq \pi$ (Whittaker and Watson, 1958), namely,

\begin{equation}\label{specialcase3_4}
	G(x)=x\tanh(x)=\frac{x^2}{\lambda_p}
\end{equation}

(\ref{specialcase3_4}) shows that $t_{sp}$ are the roots of Curie-Weiss equation that $\tanh(x)=x/\lambda_p$, namely,

\begin{equation}\label{specialcase3_5}
	t_{sp}=\frac{\sqrt{2}}{2}+\frac{\sqrt{2}}{4\lambda_p} W\left(-4\lambda_pe^{-2\lambda_p};e^{-2\lambda_p}\right)
\end{equation}
where $W(x;r)$ is the r-Lambert function (Mez$\ddot{\mathrm{o}}$ and Keady, 2016). 

From (\ref{specialcase3_1}), (\ref{specialcase3_5}) and that $\tanh(x)=x/\lambda_p$, we have,

\begin{equation}\label{specialcase3_6}
	\left\{ \begin{array}{c}
		f\left( \lambda _p,t_{sp} \right) =-\frac{1}{2}\lambda _pt_{sp}^{2}+\frac{2}{\pi}\int_0^{\pi /2}{-\frac{1}{2}\log \left( 1-2t_{sp}^{2}\sin ^2\xi \right) \mathrm{d}\xi}=-\frac{1}{2}\lambda _pt_{sp}^{2}-\log \frac{1+\sqrt{1-2t_{sp}^{2}}}{2}\\
		\left. \frac{\mathrm{d}^2f}{\mathrm{d}t^2} \right|_{t=t_{sp}}=-\lambda _p+\frac{2}{\pi}\int_0^{\pi /2}{\left( \sqrt{2}\lambda _p\sin \xi \right) ^2\left[ 1-\frac{\left( \sqrt{2}\lambda _pt_{sp}\sin \xi \right) ^2}{\lambda _{p}^{2}} \right] \mathrm{d}\xi =-\lambda _p+\lambda _{p}^{2}\left( 1-\frac{3}{2}t_{sp}^{2} \right)}\\
	\end{array} \right. 
\end{equation}

Now (\ref{eq19_2023}) reads,

\begin{align}\label{specialcase3_7}
	\begin{aligned}
		\frac{\log \mathcal{Z}}{N}&=\frac{\log 2}{N}+\left\{ -\frac{1}{2}\lambda _1t_{s1}^{2}+\log \left[\cosh \left(t_{s1}\lambda_1\right) \right] \right\} -\frac{1}{2N}\log \left| 1-\lambda _1{\rm sech}^2\left( \lambda _1t_{s1} \right) \right| \\
		&\hspace{2em} +\frac{(N-1)\log 2}{N}+\sum_{p=2}^N {\left(-\frac{1}{2}\lambda _pt_{sp}^{2}-\log \frac{1+\sqrt{1-2t_{sp}^{2}}}{2} \right)}\\ &\hspace{2em}+\frac{1}{N}\sum_{p=2}^{N}{\left[-\frac{1}{2}\log \left| -1+\lambda _p\left( 1-\frac{3}{2}t_{sp}^{2} \right) \right|\right]}
	\end{aligned}
\end{align}
where $t_{s1}=1+\frac{W\left(-4\lambda _1e^{-2\lambda _1}; e^{-2\lambda _1  }\right)}{2\lambda _1}$, $t_{sp}$ are from (\ref{specialcase3_5}). 

When $N\rightarrow \infty$, (\ref{specialcase3_7}) reads (For the sake of space we take 1D model for an example),

\begin{align}\label{specialcase3_10}
	\begin{aligned}
		-\frac{\psi}{kT}&=-zt_{s1}^{2}+\log \left[\cosh \left( 2zt_{s1} \right) \right] + \log 2 \\
		&\hspace{1em} +\sum_{p=2}^\infty {\left(-\frac{1}{2}\lambda _pt_{sp}^{2}-\log \frac{1+\sqrt{1-2t_{sp}^{2}}}{2} \right)} \\
		&\hspace{1em} -\frac{1}{2}\frac{1}{2\pi}\int\limits_0^{2\pi}\log {\bigg\vert} -1+2z\cos \omega _1 \left[ 1-\frac{3}{2}\left[ t_{sp}\left(2z \cos \omega _1\right)\right]^2 \right]{\bigg\vert}\mathrm{d}\omega _1
	\end{aligned}
\end{align}

It should be pointed out that $t_{sp}(\cdot)$ in the integral above is the function of $2z\cos \omega _1$.

\section{Further approximating and generalizing of LFEPP}\label{approx_general}

\subsection{A further approximate LFEPP of Ising model}\label{ApproxLFEPP}
It can be found that even the exact LFEPP of (\ref{specialcase3_10}) includes complicated integral because of $\mathcal{I}(\lambda_p, t)$. In the following a further approximate to $\mathcal{I}(\lambda_p, t)$ is taken into account, from which the equivalence of $t_{sp}(\cdot)$ in (\ref{specialcase3_10}) involves only algebraic calculations, albeit in a more complex form. In the following we focus on 2D and 3D Ising model because we will calculate their critical inverse temperatures ($z_c$) based on the further approximate LFEPP in this section. We start with (\ref{SCeq1}),

\begin{equation}\label{SCeq1}
	f\left( \lambda _p,t \right) =-\frac{1}{2}\lambda _pt^2+\frac{2}{\pi}\int_0^{\pi /2}{\log \left[ \cosh \left( \sqrt{2}\lambda _pt\sin \xi \right) \right] \mathrm{d}\xi}
\end{equation}

And we change the $\sqrt{2}\lambda _pt\sin \xi$ and the $\frac{1}{2}\lambda _pt^2$ into (\ref{SCeq2}) and (\ref{SCeq3}), respectively.

\begin{align}\label{SCeq2}
	\begin{aligned}
    \sqrt{2}\lambda _pt\sin \xi =\sqrt{2}z\,\,\eta \frac{\lambda _{p}^{\prime}}{\eta}\,\,t^{\prime}\,\,\sin \xi =\left( \sqrt{2}\,z\eta \right) \bar{\lambda}_pt^{\prime}\sin \xi =x_0\bar{\lambda}_pt^{\prime}\sin \xi 
	\end{aligned}
\end{align}

\begin{equation}\label{SCeq3}
-\frac{1}{2}\lambda _pt^2=\left( -\frac{1}{2}z\eta \right) \bar{\lambda}_pt^{\prime2}
\end{equation}
where $\eta$ is a constant to make $\bar{\lambda}_p <1$, and $\lambda_{p}^{\prime}=\lambda_p/z$. $t^\prime\in t$ and $t^\prime <1$ here because: (1) $t_{sp}\rightarrow 2\sqrt{2}/\pi$ when $z\rightarrow\infty$; (2) our problem is an asymptotic one, in which only the behavior of $t$ near $t_{sp}$ is taken into account. 

After this transform, $f\left( \lambda _p,t \right)=f\left( \bar{\lambda}_p,t^{\prime};z,\eta \right)$. Further we can expand $\log \left[ \cosh (x) \right]$ into a series in the neighborhood of $x_0$ ($0\leq \bar{\lambda}_pt^\prime \sin \xi < 1$), and obtain approximately $f\left( \bar{\lambda}_p,t^{\prime};z,\eta \right)$, then $t_{sp}^{\prime}$, then $f\left(\bar{\lambda }_p,t_{sp}^{\prime}; z,\eta \right)$ and then $f^{\prime\prime}\left( \bar{\lambda}_p,t_{sp}^\prime;z,\eta \right)$ in turn. If a 2-order Taylor series is used, they are (Details can be seen in Appendix \ref{Apend_preD}),

\begin{align}\label{SCeq5_zw}
	\begin{aligned}
		f\left( \bar{\lambda}_p,t^{\prime};z,\eta \right) \approx \left[ -\frac{1}{2}z\eta \,\bar{\lambda}_p+\frac{1}{2}F_2(x_0)x_{0}^{2}\bar{\lambda}_{p}^{2} \right] t^{\prime2}+\frac{2}{\pi}\left[F_1(x_0)x_0-2F_2(x_0)x_{0}^{2}\right] \bar{\lambda}_pt^{\prime}\\
		+F_2(x_0)x_{0}^{2}-F_1(x_0)x_0+F_0(x_0)
\end{aligned}
\end{align}

\begin{equation}\label{SCeq7_zw}
	t_{sp}^{\prime}=\frac{\frac{2}{\pi}\left[ F_1(x_0)x_0-2F_2(x_0)x_{0}^{2} \right]}{z\eta -F_2(x_0)x_{0}^{2}\bar{\lambda}_p}
\end{equation}

\begin{equation}\label{SCeq8_zw}
	f\left(\bar{\lambda }_p,t_{sp}^{\prime}; z,\eta \right) 
	 \approx \frac{4}{\pi ^2}z\left[F_1(x_0)-2F_2(x_0)x_{0} \right] ^2 \lambda^\prime _p +F_2(x_0)x_{0}^{2}-F_1(x_0)x_0+F_0(x_0)
\end{equation}

\begin{align}\label{SCeq11_zw}
	\begin{aligned}
		&f^{\prime\prime}\left( \bar{\lambda}_p,t_{sp}^\prime;z,\eta \right)
		&=z\lambda _{p}^{\prime}\left( -1+4zF_2(x_0)\frac{\lambda _{p}^{\prime}}{2} \right) 
	\end{aligned}
\end{align} 
where $F_2(x_0)=\frac{1}{2}{\rm sech}^2(x_0)$, $F_1(x_0)=\tanh(x_0)$ and $F_0(x_0)=\log[\cosh(x_0)]$. Note that $F_2(x_0)$, $F_1(x_0)$ and $F_0(x_0)$ here are not those in (\ref{SCeq16_2d_new}) and (\ref{SCeq16_new}).

$I_p$ now can be derived by inserting (\ref{SCeq8_zw})-(\ref{SCeq11_zw}) into the second row of (\ref{eq14_add}); And $\mathcal{Z}$ can be deduced further by the first row of (\ref{eq19}) with this $I_p$. Finally we get,

\begin{align}\label{SCeq13}
	\begin{aligned}
		\frac{\log \mathcal{Z}}{N}	& =\log 2 +\left\{ -\frac{1}{2}\lambda _1t_{s1}^{2}+\log \left[ \cosh \left( t_{s1}\lambda _1 \right) \right] \right\} -\frac{1}{2N}\log \left| 1-\lambda _1{\rm sech}^2\left( \lambda _1t_{s1} \right) \right|
		\\
		&\ \ \ \ \ +\sum_{p=2}^N\left\{\varPhi \lambda _{p}^{\prime}+F_2(x_0)x_{0}^{2}-F_1(x_0)x_0+F_0(x_0) \right\}\\
		&\ \ \ \ \ +\frac{1}{N}\sum_{p=2}^N{\left[ -\frac{1}{2}\log \left| -1+4zF_2(x_0)\frac{\lambda _{p}^{\prime}}{2} \right| \right]}
	\end{aligned}
\end{align}
where $t_{s1}=1+{W\left(-4\lambda _1e^{-2\lambda _1}; e^{-2\lambda _1  }\right)}/{2\lambda _1}$; $\varPhi=\frac{4}{\pi ^2}z\left[F_1(x_0)-2F_2(x_0)x_{0} \right] ^2$.

For 2D Ising model, (\ref{SCeq13}) reads when $N\rightarrow \infty$,

\begin{align}\label{SCeq16_2d}
	\begin{aligned}
		-\frac{\psi}{kT}
		& \approx\log 2 -2zt_{s1}^{2}+\log \left[ \cosh \left( 4zt_{s1} \right) \right]+\sum_{p=2}^\infty \left\{\varPhi \lambda _{p}^{\prime}+F_2(x_0)x_{0}^{2}-F_1(x_0)x_0+F_0(x_0) \right\} 
		\\
		&\hspace{2em}  -\frac{1}{2}\frac{1}{\left( 2\pi \right) ^2}\iint\limits_0^{2\pi}{\log \left| -1+4zF_2(x_0) \left( \cos \omega _1+\cos \omega _2 \right) \right|\mathrm{d}\omega _1\mathrm{d}\omega _2}
	\end{aligned}
\end{align} 

For 3D Ising model, (\ref{SCeq13}) reads when $N\rightarrow \infty$,

\begin{align}\label{SCeq16}
	\begin{aligned}
		-\frac{\psi}{kT}
		& \approx\log 2 -3zt_{s1}^{2}+\log \left[ \cosh \left( 6zt_{s1} \right) \right]+\sum_{p=2}^\infty \left\{\varPhi \lambda _{p}^{\prime}+F_2(x_0)x_{0}^{2}-F_1(x_0)x_0+F_0(x_0) \right\} 
		\\
		&\hspace{2em}  -\frac{1}{2}\frac{1}{\left( 2\pi \right) ^3}\iiint\limits_0^{2\pi}{\log \left| -1+4zF_2(x_0) \left( \cos \omega _1+\cos \omega _2+\cos \omega _3 \right) \right|\mathrm{d}\omega _1\mathrm{d}\omega _2\mathrm{d}\omega _3}   
	\end{aligned}
\end{align}

(\ref{SCeq16_2d}) and (\ref{SCeq16}) look "simple" in form, from which we can obtain (\ref{eq26}), (\ref{3D_small_z}), (\ref{Asymptotic_2D}) and (\ref{Asymptotic_3D}) easily. For that $z\rightarrow 0$ in the subsection \ref{SpecialCase1}, $F_2(x_0)\rightarrow \frac{1}{2}$, $F_1(x_0) \rightarrow 0$ and $F_0(x_0)\rightarrow 0$; While $F_2(x_0)\rightarrow 0$, $F_1(x_0)\rightarrow 1$ and $F_0(x_0)\rightarrow x_0-\log 2$ for that $z\rightarrow \infty$ in the subsection \ref{SpecialCase2} (Note that the results in Appendix \ref{ApendD} will be used). These results show that (\ref{SCeq16_2d}) and (\ref{SCeq16}) are correct.

For general $z$, there are large truncation errors if we approximate $\log[\cosh(x)]$ by a 2-order Taylor series above (See Fig. \ref{fig1} in Appendix \ref{Apend_preD}). To decrease these errors, the technique of economization of power series is an alternative (e.g., Arfken, 1970), in which $F_0(x_0)$, $F_1(x_0)$ and $F_2(x_0)$ are determined in an implicit manner. The reasonability of this technique, together with example formulae ((\ref{SCeq4_2})-(\ref{SCeq4_5})) for $F_0(x_0)$, $F_1(x_0)$ and $F_2(x_0)$, are shown in Fig. \ref{fig1} and in Appendix \ref{Apend_preD}. It can be found now that   $F_0(x_0)\neq\log[\cosh(x_0)]$, $F_1(x_0)\neq\tanh(x_0)$ and $F_2(x_0)\neq\frac{1}{2}{\rm sech}^2(x_0)$. 

With the new $F_0(x_0)$, $F_1(x_0)$ and $F_2(x_0)$, (\ref{SCeq16_2d}) now is,

\begin{align}\label{SCeq16_2d_new}
	\begin{aligned}
		-\frac{\psi}{kT}
		& \approx \log 2 -2zt_{s1}^{2}+\log \left[ \cosh \left( 4zt_{s1} \right) \right]-4\varPhi 
		\\
		&\hspace{2em}  -\frac{1}{2}\frac{1}{\left( 2\pi \right) ^2}\iint\limits_0^{2\pi}{\log \left| -1+4zF_2(x_0) \left( \cos \omega _1+\cos \omega _2 \right) \right|\mathrm{d}\omega _1\mathrm{d}\omega _2}    
	\end{aligned}
\end{align}

And (\ref{SCeq16}) is,

\begin{align}\label{SCeq16_new}
	\begin{aligned}
		-\frac{\psi}{kT}
		& \approx \log 2 -3zt_{s1}^{2}+\log \left[ \cosh \left( 6zt_{s1} \right) \right]-6\varPhi 
		\\
		&\hspace{2em}  -\frac{1}{2}\frac{1}{\left( 2\pi \right) ^3}\iiint\limits_0^{2\pi}{\log \left| -1+4zF_2(x_0) \left( \cos \omega _1+\cos \omega _2+\cos \omega _3 \right) \right|\mathrm{d}\omega _1\mathrm{d}\omega _2\mathrm{d}\omega _3}    
	\end{aligned}
\end{align}

For that $x_0=10\sqrt{2}z$ and that $N=28$, it can be inferred from Fig. \ref{fig1} that the absolute errors between (\ref{SCeq16_2d_new}) and (\ref{2DFE}) are small when $0<z\leq 0.5$ and/or $z>2.0$, but they increase when $0.5<z\leq 2.0$. The same are true for (\ref{SCeq16_new}) and (\ref{3DFE}). Such errors will decrease theoretically if a $N$-order Taylor series ($N>28$) of $\log[\cosh(x)]$ is adopted.

It can also be seen from (\ref{SCeq4_2})-(\ref{SCeq4_5}) that the new $F_0(x_0)$, $F_1(x_0)$ and $F_2(x_0)$ are very complicated, although they can be calculated analytically step by step from the Taylor series of $\log[\cosh(x)]$ at $x_0$. Currently,
we have no a good way to obtain simple $F_0(x_0)$, $F_1(x_0)$ and $F_2(x_0)$. 

\subsection{$z_c$ of the 2D and 3D model}\label{Tc2D3D}

With (\ref{SCeq16_2d_new}) and the acceptable errors above when $z\in [0,0.5]$, we can estimate the $z_c$ for 2D Ising model. It can be found from (\ref{SCeq16_2d_new}) that $z_c$ occurs when $-1+8zF_2(x_0)=0$. That is to say, $z_c$ is the intersection of two curves that $y=1/8z$ and that $y=F_2(x_0)=F_2(10\sqrt{2}z)$. Our calculation in Fig. \ref{fig2} in Appendix \ref{ApendDD} shows that $z_c\approx 0.439$ when a 28-order Taylor series of $\log[\cosh(x)]$ is used to evaluate $F_2(10\sqrt{2}z)$ from (\ref{SCeq4_2})-(\ref{SCeq4_5}), which is near $0.440$ from Onsager (1944) (namely from (\ref{Onsager1})). This consistency further validates that (\ref{SCeq16_2d_new}) is correct. Therefore, the same approach and parameters ($x_0=10\sqrt{2}z; N=28$) can be applied naturally to the 3D model. 

For 3D model, it can be found from (\ref{SCeq16_new}) that $z_c$ occurs when $-1+12zF_2(x_0)=0$. Caculating in Fig. \ref{fig2} in Appendix \ref{ApendDD} shows that $z_c\approx 0.218$, which is near $0.221$ that from a high-resolution Monte Carlo study on a finite-size simple cubic Ising model (Xu et al., 2018). Because the exact $z_c$ can be obtained theoretically from (\ref{SCeq16_new}) only when both a $N$-order Taylor series and/or a $N$-order Chebyshev series ($N\rightarrow\infty$) of $\log[\cosh(x)]$ are adopted, we infer that $z_c\approx 0.21\sim 0.22$ for the 3D Ising model. 

\subsection{The LFEPP of Ising model in a non-zero external magnetic field}\label{HS_non0h}

The method here can be easily generalized to the Ising model with a non-zero external magnetic field. For this purpose, Einstein summation convention will be mainly adopted in this section and Appendix \ref{ApendH} unless otherwise specified. For this case, the nearest-neighbor Ising model in $D$-dimensions ($D=1,2,3,...$) is defined in terms of the following Hamiltonian (e.g., Huang, 1987),  

\begin{align}\label{eq27}
	\begin{aligned}
	\mathcal{H} &=  -\frac{1}{2}\sum\limits_{i,j = 1}^N {{K_{ij}}} {s_i}{s_j}-\sum_{i}s_i \tilde{h}\\
	&=  -\frac{1}{2} {s_i} {K_{ij}} {s_j}- s_i \tilde{h_i}	
\end{aligned}
\end{align}
where, $\tilde{h}=\beta h$ and $h$ is the Zeeman energy associated with an external magnetic field in the $z$-direction. $\tilde{h}_i$ is an $N$-dimensional column vectors with that $[\tilde{h}]_i=\tilde{h}$. 

The corresponding partition function is, 

\begin{equation}\label{eq28}
\mathcal{Z}=\sum_{\left\{ s_i=\pm 1 \right\}}{\exp \left[ \frac{1}{2} s_i {K_{ij} s_j+{s_i}\tilde{h}_i} \right]}
\end{equation}

According to Appendix \ref{ApendH}, this partition function now is approximated as,

\begin{align}\label{eq_non0h}
	\begin{aligned}
		\mathcal{Z}& \sim \exp \left( -2N\alpha \right) \exp \left( -\frac{1}{2}\sum_p{\frac{\tilde{h}^2}{\lambda _p}} \right) 
		\\
		&\hspace{2em}  \times 2\exp \left\{ N\left[ -\frac{1}{2}\lambda _1t_{s1}^{2}+\tilde{h}t_{s1}+\log \left( \cosh \left( 2t_{s1}\frac{\lambda _1}{2} \right) \right) \right] \right\} \left| 1-\lambda _1{\rm sech}^2\left( \lambda _1t_{s1} \right) \right|^{-1/2}
		\\
		&\hspace{2em}  \times \prod_{p=2}{ 2 \exp[Nf(\lambda _p,t_{sp})]    \left| -1+\frac{\mathcal{I}^{''}\left(\lambda _p,t_{sp} \right)}{\lambda_p} \right|^{-1/2}}
	\end{aligned}
\end{align}
where details of $\alpha$, $f(\lambda _p,t)$, $\mathcal{I}^{''}\left(\lambda _p,t \right)$, $t_{s1}$, $t_{sp}$ are in Appendix \ref{ApendH} and Appendix \ref{ApendB}. Note that $t_{s1}$ is now from the equation that $t=\tanh \left( \lambda _1t \right) +\tilde{h}/\lambda _1$.

It can be found that (\ref{eq_non0h}) has three parts which are related to the non-zero external magnetic field $h$, namely, $t_{s1}$, $\tilde{h}t_{s1}$, and $\exp \left( -\frac{1}{2}\sum_p{{\tilde{h}^2}/{\lambda _p}} \right)$, and the rest of calculation is similar to Section \ref{HS_0h} and Appendix \ref{ApendB}. This shows that the method here can be easily generalized to the Ising model with a non-zero external magnetic field $h$. For brevity, we just list the corresponding results as follows,

\begin{align}\label{eq_non0h_add4}
	\begin{aligned}
		\frac{\log \mathcal{Z}}{N}&=-2\alpha -\frac{1}{2}\frac{1}{N}\sum_{p=1}^{\infty}{\frac{\tilde{h}^2}{\lambda _p}}+\left\{ -\frac{1}{2}\lambda _1t_{s1}^{2}+\tilde{h}t_{s1}+\log \left[\cosh \left( 2t_{s1}\frac{\lambda _1}{2} \right) \right] \right\} 
		\\
		&\hspace{2em}   -\frac{1}{2N}\log \left| 1-\lambda _1{\rm sech} ^2\left( \lambda _1t_{s1} \right) \right| +\log 2 +\sum_{p=2}^N f(\lambda _p,t_{sp})
		\\
		&\hspace{2em}  +\frac{1}{N}\sum_{p=2}^{N}{\left[ -\frac{1}{2}\log \left| -1+\frac{\mathcal{I}^{''}\left(2\frac{\lambda _p}{2} ,t_{sp} \right)}{2\frac{\lambda _p}{2}} \right| \right]}
	\end{aligned}
\end{align}

For 1D Ising model when $N\rightarrow\infty$, (\ref{eq_non0h_add4}) reads,

\begin{align}\label{eq_non0h_add5}
	\begin{aligned}
		-\frac{\psi}{kT}&=-2\alpha -\left( \alpha +z \right) t_{s1}^{2}+\tilde{h}t_{s1}+\log \left( \cosh \left( 2t_{s1}\left( \alpha +z \right) \right) \right) 
		\\
		&\hspace{2em} +\log 2+\sum_{p=2}^\infty f(\lambda _p,t_{sp}) -\frac{1}{2}\frac{1}{2\pi}\int_0^{2\pi}{\frac{\tilde{h}^2}{2(\alpha +z\cos \omega _1)}}\mathrm{d}\omega _1 
		\\
		&\hspace{3em} -\frac{1}{2}\frac{1}{2\pi}\int_0^{2\pi}{\log \left| -1+ \frac{\mathcal{I}^{''}\left[ 2\left( \alpha +z\cos \omega _1 \right),t_{sp} \right]}{2\left( \alpha +z\cos \omega _1 \right)} \right|\mathrm{d}\omega _1}
	\end{aligned}
\end{align}

For 2D Ising model, (\ref{eq_non0h_add4}) reads,

\begin{align}\label{eq_non0h_add6}
	\begin{aligned}
		-\frac{\psi}{kT}&=-2\alpha -\left( \alpha +2z \right) t_{s1}^{2}+\tilde{h}t_{s1}+\log \left[ \cosh \left( 2t_{s1}\left( \alpha +2z \right) \right) \right] +\log 2
		\\
		&\hspace{1em} +\sum_{p=2}^\infty f(\lambda _p,t_{sp}) -\frac{1}{2}\frac{1}{\left( 2\pi \right) ^2}\iint\limits_0^{2\pi}{\frac{\tilde{h}^2}{2\alpha +2z(\cos \omega _1+\cos \omega _2)}}\mathrm{d}\omega _1\mathrm{d}\omega _2 
		\\
		&\hspace{1em} -\frac{1}{2}\frac{1}{\left( 2\pi \right) ^2}\iint\limits_0^{2\pi}{\log \left| -1+ \frac{\mathcal{I}^{''}\left[2\left( \alpha +z(\cos \omega _1+\cos \omega _2 )\right) \right]}{2\left[ \alpha +z(\cos \omega _1+\cos \omega _2) \right]} \right|\mathrm{d}\omega _1\mathrm{d}\omega _2}
	\end{aligned}
\end{align}

For 3D Ising model, (\ref{eq_non0h_add4}) reads,

\begin{align}\label{eq_non0h_add7}
	\begin{aligned}
		-\frac{\psi}{kT}&=-2\alpha -\left( \alpha +3z \right) t_{s1}^{2}+\tilde{h}t_{s1}+\log \left[ \cosh \left( 2t_{s1}\left( \alpha +3z \right) \right) \right] +\log 2
		\\
		&\hspace{1em}+\sum_{p=2}^\infty f(\lambda _p,t_{sp}) -\frac{1}{2}\frac{1}{\left( 2\pi \right) ^3}\iiint\limits_0^{2\pi}{\frac{\tilde{h}^2}{2\alpha +2z(\cos \omega _1+\cos \omega _2+\cos \omega _3)}}\mathrm{d}\omega _1\mathrm{d}\omega _2\mathrm{d}\omega _3 
		\\
		&\hspace{0.5em} -\frac{1}{2}\frac{1}{\left( 2\pi \right) ^3}\iiint\limits_0^{2\pi}{\log \left| -1+ \frac{\mathcal{I}^{''}\left[2\left( \alpha +z(\cos \omega _1+\cos \omega _2+\cos \omega _3) \right),t_{sp} \right]}{2\left( \alpha +z(\cos \omega _1+\cos \omega _2+\cos \omega _3) \right)} \right| \mathrm{d}\omega _1\mathrm{d}\omega _2\mathrm{d}\omega _3}
	\end{aligned}
\end{align}

\section{Discussions and further work}

By the classic mathematical-physical approaches, we obtain the LFEPP of $N$-D Ising model without the external magnetic field from its functional integral. The final LFEPP depends mainly on an integral including some parameters of $\mathcal{I}(\lambda_p, t)$. If $\mathcal{I}(\lambda_p, t)$ can be solved (probably using special functions), then the LFEPP can be obtained (asymptotically) exactly. For 1D-3D Ising models, the LFEPPs have similar forms and structures. So it is natural to extend LFEPPs from 1D-2D models to 3D model. This approach, using only the knowledge of mathematical analysis, linear algebra and asymptotic computation, should be more elementary than those approaches using specialized knowledge and special skills, for example, Transfer matrix method for 2D and 3D Ising model. For 1D model, however, it seems that this method is more difficult to be operated than matrix method.  

When $z\rightarrow 0$, our approximate LFEPPs for 1-2D Ising model in subsection \ref{SpecialCase1} are nearly the same to results from  (\ref{KW1}) and (\ref{Onsager1}) in the section \ref{intro} and those in Feynman (2018), except for the differences of "$\log 2$" and the minus sign before the integral representation. "$\log 2$" may be related to the normalization factor, or the reference point for the LFEPP; The minus sign can be converted into a plus sign by an appropriate transformation. When $z\rightarrow \infty$, the approximate LFEPPs for 1-2D Ising model in subsection \ref{SpecialCase2}, are of the same order with the well-known results from  (\ref{KW1}) and (\ref{Onsager1}). On the other hand, with our  approximate solution (subsection \ref{ApproxLFEPP}), we obtain semi-analytically $z_c$ ($ \approx 0.439$) of the 2D Ising model  which also agrees with the well-known result from (\ref{Onsager1}). These consistences show that our method here is reasonable, and allow us to  generalize naturally it to 3D Ising model. 

When $z\rightarrow 0$, our approximate LFEPPs for 1-2D Ising model in subsection \ref{SpecialCase1}, are nearly the same to those from Gaussian model. Gaussian model assumes that the probability of finding a give spin $s_i$ between $s_i$ and $s_i+{\rm d}s_i$ is given by $(2\pi)^{-1/2}\exp (-s^2_i/2){\rm d}s_i$ (e.g., Berlin and Kac, 1952). These two consistent results show that 3D Ising model is also Gaussian with that $\langle s_i\rangle =0$ and that $\langle s_i^2\rangle =1$ when $z$ is small.    

Furthermore, we generalize this method to Ising model with an external magnetic field. It can be found that only three more items, which are related to this non-zero external magnetic field, are included in the partition function additionally. Although we have to introduce a parameter $2\alpha$ to make the $f(\lambda_p/2)$ be regular in Appendix \ref{ApendC}, and the form of the LFEPP is more complicated, we can obtain the high dimensional ($>=2$) LFEPP Ising model with an external magnetic field naturally. Expect that $\alpha > \lambda _{p_{\rm max}}$ (where $\lambda _{p_{\rm max}}$ is the maximum eigenvalue of $K_{ij}$), we still have no idea on what $2\alpha$ should be, although it seems and should have no effect on the final LFEPP.   

Some new difficulties arise in our approach. For example, how to obtain a simple expression of $\mathcal{I}(\lambda_p, t)$, then $t_{sp}$, and then $\sum_{p=2}^\infty {f(\lambda_p,t_{sp})}$? Although we suggest an approximate solution (subsection \ref{ApproxLFEPP}) for analytically calculating, it is not only complicated in form except for two limiting cases, but also has a low precision in calculating $z_c$. How to obtain an elementary or simple $\mathcal{I}(\lambda_p, t)$ (if possible) is the main work in future.

\vspace{7em}

{\Large\bf  Acknowledges}

\vspace{2em}

This research is funded by National Natural Science Foundation of China (Grant No. 42374109). 

\vspace{2em}

\appendix
\renewcommand{\theequation}{\thesection.\arabic{equation}} 
\setcounter{equation}{0}
\section{Matrix $K_{ij}$}\label{ApendA}

For 1-dimensional (1D) Ising model, let $\mathrm{\bf K=A}$, and $\mathrm{\bf A}$ is an $N\times N$ matrix (e.g., Dixon et al., 2001). $\mathrm{\bf A}$ is a circulant matrix when an appropriate periodicity is used on the boundary of the Ising model.

\[{\bf A} =z\times \left[ {\begin{array}{*{20}{c}}
		0&1&0&0& \cdot & \cdot & \cdot &1\\
		1&0&1&0&{}&{}&{}&{}\\
		0&1&0&1&{}&{}&{}&{}\\
		0&0&1&0&{}&{}&{}&\cdot\\
		\cdot &\cdot&\cdot&{}&{}&{}&{}&\cdot\\
		\cdot &\cdot&\cdot&{}&{}&{}&{}&\cdot\\
		\cdot &\cdot&\cdot&{}&{}&{}&{}&1\\
		1&0&0&{}&{}&{}&1&0
\end{array}} \right]\]
where $z=\frac{\epsilon}{kT}$, $\epsilon$ the interaction energy, $T$ the temperature, and $k$ Boltzmann constant. The $p$-th ($p=1,2,\cdots $) eigenvalue of $\mathrm{\bf A}$ is (e.g., Berlin and Kac, 1952),

\begin{equation}\label{eig_1D}
\lambda _p=2z \cos \frac{2\pi}{N}\left( p-1 \right) 
\end{equation}

For 2-dimensional (2D) Ising model, let $\mathrm{\bf K=B}$, and $\mathrm{\bf B}$ can be constructed from $\mathrm{\bf A}$ like the following (e.g., Dixon et al., 2001),

\[{\bf B} = \left[ {\begin{array}{*{20}{c}}
		{\bf A}&{\bf I}&0&0& \cdot & \cdot & \cdot &{\bf I}\\
		{\bf I}&{\bf A}&{\bf I}&0& \cdot & \cdot & \cdot &0\\
		0&{\bf I}&{\bf A}&{\bf I}& \cdot & \cdot & \cdot &{}\\
		0&0&{\bf I}&{}&{}&{}&{}&{\bf I}\\
		{\bf I}&0& \cdot & \cdot & \cdot &{}&{\bf I}&{\bf A}
\end{array}} \right]\]

Generally, 2D Ising model is on a rectangular lattice with $n_1$ sites in a row, $n_2$ rows, so that $N=n_1 n_2$. $\mathrm{\bf B}$ is also a circulant matrix when an appropriate periodicity is used on the boundary. Its $p$-th eigenvalue is (e.g., Berlin and Kac, 1952),

\begin{equation}\label{eig_2D}
\lambda _p=2z \cos \frac{2\pi}{N}\left( p-1 \right) +2z \cos \frac{2\pi n_1}{N}\left( p-1 \right) 
\end{equation}

For 3-dimensional (3D) Ising model, let $\mathrm{\bf K=C}$, and $\mathrm{\bf C}$ can be constructed from $\mathrm{\bf B}$ like the following (e.g., Dixon et al., 2001),

\[\bf C = \left[ {\begin{array}{*{20}{c}}
		{\bf B}&{\bf I}&{\bf O}& \cdots & \cdots &{}&{}&{\bf I}\\
		{\bf I}&{\bf B}&{\bf I}&{}&{}&{}&{}&{\bf O}\\
		{\bf O}&{\bf I}&{\bf B}&{}&{}&{}&{}&{}\\
		{}&{}&{}&{}&{}&{}&{}&{\bf O}\\
		{\bf O}&{}&{}&{}&{}&{}&{}&{\bf I}\\
		{\bf I}&{\bf O}& \cdots & \cdots &{\bf O}&{}&{}&{\bf B}
\end{array}} \right]\]

Generally, 3D Ising model is on a lattice with $n_1$ sites in a row, $n_2$ rows in a plane, and $n_3$ planes so that the total number of sites $N=n_1 n_2 n_3$. $\mathrm{\bf C}$ is also a circulant matrix when an appropriate periodicity is used on the boundary. Its $p$-th eigenvalue is (e.g., Berlin and Kac, 1952),

\begin{equation}\label{eig_3D}
\lambda _p=2z \cos \frac{2\pi}{N}\left( p-1 \right) +2z \cos \frac{2\pi n_1}{N}\left( p-1 \right) +2z \cos \frac{2\pi n_1n_2}{N}\left( p-1 \right) 
\end{equation}

For convenience, we define that $\lambda_p^\prime=\frac{\lambda_p}{z}$, and,

 \begin{equation}\label{lambdapp}
 \lambda_p^\prime=
 \begin{cases}
 	2\cos \frac{2\pi}{N}\left( p-1 \right)&		1\mathrm{D}\\
 	2\cos \frac{2\pi}{N}\left( p-1 \right) +2\cos \frac{2\pi n_1}{N}\left( p-1 \right)&		2\mathrm{D}\\
 	2\cos \frac{2\pi}{N}\left( p-1 \right) +2\cos \frac{2\pi n_1}{N}\left( p-1 \right) +2\cos \frac{2\pi n_1n_2}{N}\left( p-1 \right)&		3\mathrm{D}\\
 \end{cases}
\end{equation}

Generally, we assume that $\mathbf{K}$ is positive defined (when $N\rightarrow \infty$), as the previous scientists did (e.g., Ginzburg and Landau, 1950; Amit and Martin-Mayor, 2005). However, from (\ref{eig_1D}) to (\ref{eig_3D}), it can be found that the eigenvalues may be zero or negative when $N$ is finite. If this happens, to make $\mathbf{K}$ in the Hubbard-Stratonovich transformation be positive, we may write,

\begin{align}\label{bar_K}
	\begin{aligned}
	\sum_{ij}{K_{ij}s_is_j}&=N2\alpha -N2\alpha +\sum_{ij}{K_{ij}s_is_j}\\
	&=-N2\alpha +2\alpha \sum_i{s_{i}^{2}}+\sum_{ij}{K_{ij}s_is_j}
	\\
	&=-N2\alpha +\sum_{ij}{\widetilde{K}_{ij}s_is_j}
	\end{aligned}
\end{align}
where $\alpha$ is a real positive to make the eigenvalues of $\mathbf{\widetilde{K}}$ are positive. $\mathbf{\widetilde{K}}$ is not a circulant matrix now but a Toeplitz one, whose diagonal elements are $2\alpha$  (The original is zero) . Since $N$ is very large, $\mathbf{\widetilde{K}}$ is asymptotically equivalent to a circulant matrix according to Szeg$\ddot{o}$ theorem.  $\mathbf{\widetilde{K}}$ has eigenvalues as follows,

For 1D Ising model,

\begin{equation}\label{eig_1D_a}
\lambda _p=2\alpha+2z \cos \frac{2\pi}{N}\left( p-1 \right) 
\end{equation}

For 2D Ising model,
\begin{equation}\label{eig_2D_a}
\lambda _p=2\alpha+2z \cos \frac{2\pi}{N}\left( p-1 \right) +2z \cos \frac{2\pi n_1}{N}\left( p-1 \right) 
\end{equation}

For 3D Ising model,
\begin{equation}\label{eig_3D_a}
\lambda _p=2\alpha+2z \cos \frac{2\pi}{N}\left( p-1 \right) +2z \cos \frac{2\pi n_1}{N}\left( p-1 \right) +2z \cos \frac{2\pi n_1n_2}{N}\left( p-1 \right) 
\end{equation}

And for any cases above, the corresponding real, orthogonal eigenvectors of $K_{ij}$, normalized to unity, are given by,

\begin{equation}\label{eig_vector}
V_{ij}=N^{-1/2}\left[ \cos \frac{2\pi}{N}\left( i-1 \right) \left( j-1 \right) +\sin \frac{2\pi}{N}\left( i-1 \right) \left( j-1 \right) \right] 
\end{equation}

\setcounter{equation}{0}

\section{Calculating $I_p$}\label{Cal_Ip}\label{ApendB}

In this section, we focus on the following integral,

\begin{equation}\label{eq9}
	I_p=
	{\int_{-\infty}^{+\infty}{\exp \left( -\frac{1}{2}\lambda _py_{p}^{2} \right) \times \exp \left\{ \sum_i{\log \left[\cosh \left( V_{ip}\lambda _py_p \right) \right]} \right\} }}\mathrm{d}y_p
\end{equation}

where 
\begin{align}
	\begin{aligned}
		V_{ip}&	=N^{-1/2}\left[ \cos \frac{2\pi}{N}\left( i-1 \right) \left( p-1 \right) +\sin \frac{2\pi}{N}\left( i-1 \right) \left( p-1 \right) \right]\\
		&=N^{-1/2}\overline{V}_{ip}
	\end{aligned}
\end{align}

When $p>1$,

\begin{align}\label{eq10}
	\begin{aligned}
		I_p &=\int_{-\infty}^{+\infty}{\exp \left( -\frac{1}{2}\lambda _py_{p}^{2} \right)} \exp \left\{ \sum_i{\log \left[\cosh \left( V_{ip}\lambda _py_p \right) \right]} \right\}\mathrm{d}y_p \\
		&\xlongequal{y_p=yN^{-1/2}}  \int_{-\infty}^{+\infty}{\exp \left( -\frac{1}{2}\lambda _p\frac{y^2}{N} \right)  \exp \left\{ \sum_i{\log \left[ \cosh \left( \frac{\overline{V}_{ip}\lambda _py}{N} \right) \right]} \right\} \mathrm{d}\frac{y}{N^{1/2}}}\\
		&\xlongequal{y=tN} N^{1/2}\int_{-\infty}^{+\infty}{\exp \left( -\frac{1}{2}\lambda _pNt^2 \right)}\\
		&\hspace{2em} \times \exp \left\{ \sum_{i}{\log \left[ \cosh \left[ \left( \cos \frac{2\pi}{N}\left( i-1 \right) \left( p-1 \right) +\sin \frac{2\pi}{N}\left( i-1 \right) \left( p-1 \right) \right) \lambda _pt \right] \right]} \right\}\mathrm{d}t	
	\end{aligned}
\end{align}

Next the following sum is taken into account,

\begin{equation}\label{eq11}
	J=\sum_{i}{\log \left[ \cosh \left[ \left( \cos \frac{2\pi}{N}\left( i-1 \right) \left( p-1 \right) +\sin \frac{2\pi}{N}\left( i-1 \right) \left( p-1 \right) \right) \lambda _pt \right] \right]}
\end{equation}

It can be found that,

\begin{equation}\label{eq12}
	\frac{J}{N}=\frac{\log[\cosh(\lambda_pt)]+\sum_{i=2} {\log \left[ \cosh \left[ \left( \cos \frac{2\pi}{N}\left( i-1 \right) \left( p-1 \right) +\sin \frac{2\pi}{N}\left( i-1 \right) \left( p-1 \right) \right) \lambda _pt \right] \right]}}{N}
\end{equation}

Because $\log[\cosh(\lambda_pt)]$ is finite when $\lambda_pt < \infty$ (The case that $\lambda_pt \rightarrow \infty$ will be taken into account in Appendix \ref{ApendE}), $\frac{\log[\cosh(\lambda_pt)]}{N}\rightarrow 0$ when $N\rightarrow \infty$. Thus,

\begin{equation}\label{eq12+1}
	\frac{J}{N}=\frac{\sum_{i=2}{\log \left[ \cosh \left[ \left( \cos \frac{2\pi}{N}\left( i-1 \right) \left( p-1 \right) +\sin \frac{2\pi}{N}\left( i-1 \right) \left( p-1 \right) \right) \lambda _pt \right] \right]}}{N}
\end{equation}

Subdivide the interval $0-2\pi$ into $N$ equal intervals of length $2\pi/N$ ($=\Delta\theta$), and let $\theta =(i-1)\Delta\theta$. When $\Delta\theta\rightarrow 0$, we have,

\begin{align}\label{eq13}
	\begin{aligned}
		\frac{J}{N}&=\frac{1}{2\pi}\sum_{\theta =\Delta \theta}^{2\pi -\Delta \theta}{\log \left[ \cosh \left[ \lambda _pt\left( \cos \left( p-1 \right) \theta +\sin \left( p-1 \right) \theta \right) \right] \right]}\Delta \theta\\
		&\rightarrow\frac{1}{2\pi}\int_0^{2\pi}{\log \left[ \cosh \left[ \lambda _pt\left( \cos \left( p-1 \right) \theta +\sin \left( p-1 \right) \theta \right) \right] \right] \mathrm{d}\theta}\\
		&\xlongequal{(p-1)\theta =\xi}\frac{1}{2\pi \left( p-1 \right)}\int_0^{2\left( p-1 \right) \pi}{\log \left[ \cosh \left[ \lambda _pt\left( \cos \xi +\sin \xi \right) \right] \right] \mathrm{d}\xi}\\
		&=\frac{1}{2\pi}\int_0^{2\pi}{\log\left[\cosh\left[ \lambda _pt\left(\cos \xi +\sin \xi \right)\right]\right] \mathrm{d}\xi}\\
		& =\frac{2}{\pi}\int_{0}^{\pi /2}{\log \left[ \cosh \left( \sqrt{2}\lambda _pt\sin \xi \right) \right] \mathrm{d}\xi}\\
		&=\frac{2\sqrt{2}\lambda_p t}{\pi}-\log2+\frac{2}{\pi}\int_{0}^{\pi /2}{\log \left( 1+ e^{-2\sqrt{2}\lambda _pt\sin \xi} \right) \mathrm{d}\xi}
	\end{aligned}
\end{align}

Therefore insert $J$ into (\ref{eq10}), and we obtain,   

\begin{align}\label{eq14}
	\begin{aligned}
		I_p	&=N^{1/2}\int_{-\infty}^{+\infty}  \exp \left\{ -\frac{1}{2}\lambda_p Nt^2+N\frac{2}{\pi} \int_0^{\pi /2}{\log \left[ \cosh \left( \sqrt{2}\lambda_pt\sin \xi \right) \right] \mathrm{d}\xi}  \right\}  \mathrm{d}t\\
		&	=N^{1/2}\int_{-\infty}^{+\infty}{\exp \left[ Nf\left( \lambda _p,t \right) \right] \mathrm{d}t}
	\end{aligned}
\end{align}

where $f\left( \lambda _p,t \right)$ is,

\begin{equation}\label{eq15}
	f\left( \lambda _p,t \right) =-\frac{1}{2}\lambda _pt^2+\frac{2}{\pi} \int_0^{\pi /2}{\log \left[ \cosh \left( \sqrt{2}\lambda _pt\sin \xi \right) \right] \mathrm{d}\xi}=-\frac{1}{2}\lambda _pt^2+\mathcal{I}(\lambda_p, t)
\end{equation}

$I_p$ can be solved by Laplace's asymptotic method. The required ${\rm d}f /{\rm d}t$ and ${\rm d}^2 f/{\rm d} t^2$ are as follows,

\begin{equation}\label{eq16}
	\frac{{\rm d} f}{{\rm d} t}=-\lambda _pt+\mathcal{I}'(\lambda_p, t)
\end{equation}  

\begin{equation}\label{eq17}
	\frac{{\rm d} ^2f}{{\rm d} t^2}=f^{''}(\lambda_p, t)=-\lambda _p+\mathcal{I}{''}(\lambda_p, t)
\end{equation}

Therefore,

\begin{align}\label{eq14_add}
	\begin{aligned}
		I_p	&= N^{1/2}\int_{-\infty}^{+\infty}{\exp \left[ Nf\left( \lambda _p,t \right) \right] \mathrm{d}t}\\
		&\sim N^{1/2} {\ 2\ } \exp \left[ Nf(\lambda_p,t_{sp})\right] \sqrt{\frac{2\pi}{N\left| f^{''}\left( \lambda _p,t_{sp} \right) \right|}}\\
		&= \left\{
		\begin{array}{ll}
			2 \exp \left[ Nf(\lambda_p,t_{sp})\right] \sqrt{\frac{2\pi}{\left| -\lambda _p+\mathcal{I} ^{''}\left( \lambda _p,t_{sp} \right) \right|}}& t_{sp}\neq 0\\
			2 \exp \left[ Nf(\lambda_p,0)\right] \sqrt{\frac{\pi}{2 \left| -\lambda _p+\mathcal{I} ^{''}\left( \lambda _p,0 \right) \right|}}& t_{sp}=0
			\end{array}
			\right.			
	\end{aligned}
\end{align}
where $t_{sp}$ is from that $\frac{{\rm d} f}{{\rm d} t}=0$ (Eq. (\ref{eq16})), namely that $t=\mathcal{I}'(\lambda_p, t)/\lambda_p$. 

When $p=1$, it can be found from (\ref{eq10}) that,

\begin{align}\label{eq18_appendB}
	\begin{aligned}
		I_1	&=N^{1/2}\int_{-\infty}^{+\infty}\exp \left\{ N\left[ -\frac{1}{2}\lambda _1t^2+\log \left( \cosh \left( \lambda _1t \right) \right) \right] \right\} \mathrm{d}t\\
		&\sim \left\{
		\begin{array}{ll}
			2\exp \left\{ N\left[ -\frac{1}{2}\lambda _1 t_{s_1}^{2}+\log \left( \cosh \left( \lambda _1 t_{s_1} \right) \right) \right] \right\} \sqrt{\frac{2\pi}{\vert  -\lambda _1+\lambda _{1}^{2}{\rm sech} ^2\left( \lambda _1 t_{s_1} \right) \vert }}& t_{s1}\neq 0\\
			2 \sqrt{\frac{\pi}{2\vert  -\lambda _1+\lambda _{1}^{2} \vert }}& t_{s1}=0
		\end{array}
		\right.		
	\end{aligned}
\end{align}
where $I_1$ is solved by Laplace's asymptotic method. $t_{s_1}$ is from equation that $t=\tanh(\lambda_1 t)$. $t_{s1}=1+{W\left(-4\lambda _1e^{-2\lambda _1}; e^{-2\lambda _1  }\right)}/{2\lambda _1}$, where $W(x;r)$ is the r-Lambert function (Mez$\ddot{\mathrm{o}}$ and Keady, 2016).

\setcounter{equation}{0} 
\section{On $N^{-1}\sum_{p=1}^N{f\left( \frac{\lambda _p}{2} \right)}$ when $N\rightarrow \infty$}\label{ApendC}

It should be pointed out firstly that we mainly prove $S(f)$ for 3D Ising model. Those for 1D and 2D Ising model are from Berlin and Kac (1952) with minor modification.

The following sum is taken into account,

\[
S\left( f \right) =\lim_{N\rightarrow \infty} N^{-1}\sum_{p=1}^N{f\left( \frac{\lambda _p}{2} \right)}
\]

The largest eigenvalue occurs for that $p=1$, and we assume that $f(t)$ is regular when $t>\frac{\lambda _1}{2}$. Consequently,

\[
N^{-1}\sum_{p=1}^N{f\left( \frac{\lambda _p}{2} \right)}=N^{-1}f\left( \frac{\lambda _1}{2} \right) +N^{-1}\sum_{p=2}^N{f\left( \frac{\lambda _p}{2} \right)}
\]

If $f(\frac{\lambda_1}{2})$ is finite,

\begin{equation*}
	S\left( f \right) =\lim_{N\rightarrow \infty} N^{-1}\sum_{p=2}^N{f\left( \frac{\lambda _p}{2} \right)}
\end{equation*}

For {\bf 1D Ising model}, we have,

\begin{equation*}
	\frac{\lambda _p}{2}=z \cos \frac{2\pi}{N}\left( p-1 \right) 
\end{equation*}
where we assume that $\alpha=0$ for simplicity. $p=2,3,...$

Subdivide the interval $0-2\pi$ into $N$ equal intervals of length $\frac{2\pi}{N}$ ($=\Delta\omega_1$), and let $\omega_1=(p-1)\Delta\omega_1$. When $\Delta\omega\rightarrow 0$, we have,

\begin{equation*}
	N^{-1}\sum_{p=2}^N{f\left( \frac{\lambda _p}{2} \right)}=\frac{1}{2\pi}\sum_{\omega _1=\Delta \omega _1}^{2\pi -\Delta \omega _1}{f\left(z\cos \omega _1 \right)}\Delta \omega _1
\end{equation*}

\begin{align}\label{Sf_1D}
	\begin{aligned}
		S\left( f \right)&=\lim_{\Delta \omega _1\rightarrow 0} \frac{1}{2\pi}\sum_{\omega _1=\Delta \omega _1}^{2\pi -\varDelta \omega _1}{f\left( z\cos \omega _1 \right)}\varDelta \omega _1\\
		&=\frac{1}{2\pi}\int_0^{2\pi}{f\left( z\cos \omega _1 \right) \mathrm{d}\omega_1}
	\end{aligned}
\end{align}

For {\bf 2D Ising model},

$$
\frac{\lambda _p}{2}=z \cos \frac{2\pi}{N}\left( p-1 \right) +z\cos \frac{2\pi n_1}{N}\left( p-1 \right) 
$$
where $N=n_1 n_2$.

Let $p-1=p_2+p_1n_2$, $p_1=0,1,...n_1-1$ and $p_2=0,1,2,...n_2-1$, then

$$
\sum_{p=2}^N{ }=\sum_{p_1=0}^{n_1-1}{ }\sum_{p_2=1}^{n_2-1}{ }
$$

and,

$$
\frac{\lambda _p}{2}=z\cos \left( \frac{2\pi}{n_1n_2}p_2+\frac{2\pi}{n_1}p_1 \right) +z\cos \left( \frac{2\pi}{n_2}p_2+2\pi p_1 \right) 
$$

Let $\Delta \omega _2=\frac{2\pi}{n_2}$, $\omega _2=\frac{2\pi}{n_2}p_2 \in [0,2\pi ]$. Since $n_1,n_2$ become large, $\frac{2\pi}{n_1n_2}p_2=\frac{\omega _2}{n_1}\rightarrow 0
$

and,

$$
\frac{\lambda _p}{2}=z\cos \left( \frac{2\pi}{n_1}p_1 \right) +z\cos \omega _2\mathrm{  }
$$

Furthermore,

$$
\sum_{p_2=0}^{n_2-1}{=}\frac{n_2}{2\pi}\sum_{2{{\pi}/{n_2}}}^{2\pi -2{{\pi}/{n_2}}}{\varDelta \omega _2}\rightarrow \frac{n_2}{2\pi}\int_0^{2\pi}{\mathrm{d}\omega _2}
$$

Hence,

$$
N^{-1}\sum_{p=2}^N{f\left( \frac{\lambda _p}{2} \right)}\rightarrow \frac{n_2}{2\pi N}\sum_{p_1=0}^{n_1-1}{\int_0^{2\pi}{f\left( \frac{\lambda _p}{2} \right) \mathrm{d}\omega _2}}
$$

Let $\Delta \omega _1=2\pi /n_1$, $\omega _1=\left( 2\pi /n_1 \right) p_1$, then,

$$
\frac{\lambda _p}{2}=z\cos \omega _1+z\cos \omega _2
$$

\begin{align*}
	\begin{aligned}
		N^{-1}\sum_{p=2}^N{f\left( \frac{\lambda _p}{2} \right)}\rightarrow \frac{n_2}{2\pi N}\frac{n_1}{2\pi}\sum_{2{{\pi}/{n_1}}}^{2\pi -2{{\pi}/{n_1}}}{\Delta \omega _1\int_0^{2\pi}{f\left( \frac{\lambda _p}{2} \right) \mathrm{d}\omega _2}}\\
		\rightarrow \frac{1}{\left( 2\pi \right) ^2}\int_0^{2\pi}{\int_0^{2\pi}{f\left( \frac{\lambda _p}{2} \right) \mathrm{d}\omega _1\mathrm{d}\omega _2}}\hspace{3.2em}\\
	\end{aligned}
\end{align*}

Finally,

\begin{equation}\label{Sf_2D}
	S\left( f \right) =\frac{1}{\left( 2\pi \right) ^2}\int_0^{2\pi}{\int_0^{2\pi}{f\left(z\cos \omega _1+z \cos \omega _2 \right) \mathrm{d}\omega _1\mathrm{d}\omega _2}}
\end{equation}

For {\bf 3D Ising model},

\[
\frac{\lambda _p}{2}=z \cos \frac{2\pi}{N}\left( p-1 \right) +z \cos \frac{2\pi n_1}{N}\left( p-1 \right) +z \cos \frac{2\pi n_1n_2}{N}\left( p-1 \right) 
\]
where $N=n_1n_2n_3$.

Let $p-1=rn_3+q$, $r=0,1,\cdots n_1n_2-1$, and $q=0,1,\cdots n_3-1$, then,

\[
\sum_{p=2}^N{}=\sum_{r=0}^{n_1n_2-1}{}\sum_{q=1}^{n_3-1}{}
\]

and
$$
\frac{\lambda _p}{2}=z \cos \left( \frac{2\pi r}{n_1n_2}+\frac{2\pi q}{n_1n_2n_3} \right) +z \cos \left( \frac{2\pi r}{n_2}+\frac{2\pi q}{n_2n_3} \right) +z \cos \left( 2\pi r+\frac{2\pi q}{n_3} \right) 
$$

Let $\Delta \omega _3=2\pi /n_3$, $\omega _3=2\pi q/n_3\in [0,2\pi]$. Since $r$ is an integer, $\cos \left( 2\pi r+2\pi q/n_3 \right) =\cos \omega _3$. As $n_1,n_2,n_3$ become large, we have,

$$
\frac{2\pi q}{n_2n_3}=\frac{\omega _3}{n_2}\rightarrow 0
$$

$$
\frac{2\pi q}{n_1n_2n_3}=\frac{\omega _3}{n_1n_2}\rightarrow 0
$$

$$
\frac{\lambda _p}{2}=z \cos \left( \frac{2\pi r}{n_1n_2} \right) +z \cos \left( \frac{2\pi r}{n_2} \right) +z \cos \omega _3
$$

Furthermore,

$$
\sum_{q=0}^{n_3-1}{}=\frac{n_3}{2\pi}\sum_{2\pi /n_3}^{2\pi -2\pi /n_3}{\Delta \omega _3}\rightarrow \frac{n_3}{2\pi}\int_0^{2\pi}{\mathrm{d}\omega _3}
$$

Hence,

$$
N^{-1}\sum_{p=2}^N{f\left( \frac{\lambda _p}{2} \right)}\rightarrow \frac{n_3}{2\pi N}\int_0^{2\pi}{\mathrm{d}\omega _3\sum_{r=0}^{n_1n_2-1}{f\left( \frac{\lambda _p}{2} \right)}}
$$

Let $r=tn_2+s$, $t=0,1,\cdots ,n_1-1$, $s=0,1,\cdots ,n_2-1$,  

$$
\frac{\lambda _p}{2}=z \cos \left( \frac{2\pi t}{n_1}+\frac{2\pi s}{n_1n_2} \right) +z \cos \left( 2\pi t+\frac{2\pi s}{n_2} \right) +z \cos \omega _3
$$

Let $\Delta \omega _2=2\pi /n_2$, $\omega _2=2\pi s/n_2$, then  $\cos \left( 2\pi t+2\pi s/n_2 \right) =\cos \omega _2$. As $n_1\rightarrow \infty$, $\frac{2\pi s}{n_1n_2}=\frac{\omega _2}{n_1}\rightarrow 0$, we have,

\[
\frac{\lambda _p}{2}=z \cos \left( \frac{2\pi t}{n_1} \right) +z \cos \omega _2+ z\cos \omega _3
\]

Furthermore,

\begin{align*}
	\begin{aligned}
		\sum_{r=0}^{n_1n_2-1}{}&=\sum_{t=0}^{n_1-1}{}\sum_{s=0}^{n_2-1}{}
		\\
		&=\sum_{t=0}^{n_1-1}{\frac{n_2}{2\pi}\sum_{2\pi /n_2}^{2\pi -2\pi /n_2}{\varDelta \omega _2}}\rightarrow \sum_{t=0}^{n_1-1}{\frac{n_2}{2\pi}\int_0^{2\pi}{\mathrm{d}\omega _2}}
	\end{aligned}
\end{align*}

Hence,

\[
N^{-1}\sum_{p=2}^N{f\left( \frac{\lambda _p}{2} \right)}\rightarrow \frac{n_2n_3}{\left( 2\pi \right) ^2N}\iint\limits_0^{2\pi}{\mathrm{d}\omega _2\mathrm{d}\omega _3}\sum_{t=0}^{n_1-1}{f\left( \frac{\lambda _p}{2} \right)}
\]

Now set $\varDelta \omega _1=2\pi /n_1$, $\omega _1=2\pi t/n_1$,

$$
N^{-1}\sum_{p=2}^N{f\left( \frac{\lambda _p}{2} \right)}\rightarrow \frac{1}{\left( 2\pi \right) ^3}\iiint\limits_0^{2\pi}{f\left( \frac{\lambda _p}{2} \right) \mathrm{d}\omega _1\mathrm{d}\omega _2\mathrm{d}\omega _3}
$$

And finally,

\begin{equation}\label{Sf_3D}
	S\left( f \right) =\frac{1}{\left( 2\pi \right) ^3}\int_0^{2\pi}{\int_0^{2\pi}{\int_0^{2\pi}{f\left(z \cos \omega _1+z \cos \omega _2+z \cos \omega _3 \right) \mathrm{d}\omega _1\mathrm{d}\omega _2}}}\mathrm{d}\omega _3
\end{equation}

\setcounter{equation}{0} 
\section{On $\sum_{p=2}^N{\varPhi \lambda _{p}^{\prime}}$}\label{ApendD}

The identity (\ref{eq_app_1}) will be used in the following,

\begin{equation}\label{eq_app_1}
\sum_{k=0}^N{\cos kx}=\frac{1}{2}\left[ 1+\frac{\sin \left( N+\frac{1}{2} \right) x}{\sin \frac{x}{2}} \right]
\end{equation}

For 1D Ising model,

\begin{align}\label{eq_app_2_1}
	\begin{aligned}	
		\sum_{p=2}^N{\varPhi \lambda _{p}^{\prime}}&=\varPhi \sum_{p=2}^N{2\cos \frac{2\pi}{N}\left( p-1 \right)}
		\\
		&\,\,  =2\varPhi \sum_{k=1}^{N-1}{\cos \frac{2\pi}{N}k}
		\\
		&\,\,  =2\varPhi \left[ \left( \sum_{k=0}^N{\cos \frac{2\pi}{N}k} \right) -\cos\frac{2\pi}{N} 0 -\cos\frac{2\pi}{N} N \right] 
		\\
		&\,\,  =2\varPhi \left\{ \frac{1}{2}\left[ 1+\frac{\sin \left( N+\frac{1}{2} \right) \frac{2\pi}{N}}{\sin \frac{2\pi}{2N}} \right] -2 \right\} 
		\\
		&\,\,   =2\varPhi \left[ \frac{1}{2}\left( 1+1 \right) -2 \right] 
		\\
		&\,\,   =-2\varPhi 	
	\end{aligned}
\end{align}
where $\varPhi$ is a constant.

For 2D Ising model ($N=n_1\times n_2$),

\begin{align}\label{eq_app_3}
	\begin{aligned}	
		\sum_{p=2}^N{\varPhi \lambda _{p}^{\prime}}	
		&\,\,  =\varPhi \sum_{p=2}^N{2\cos \frac{2\pi}{N}\left( p-1 \right) +2\cos \frac{2\pi n_1}{N}\left( p-1 \right)}
		\\
		&\,\,  =2\varPhi \sum_{k=1}^{N-1}{\cos \frac{2\pi}{N}k}+\cos \frac{2\pi}{n_2}k\\
		&\,\,  =2\varPhi \left[ \left( \sum_{k=0}^N{\cos \frac{2\pi}{N}k} \right) -2+\left( \sum_{k=0}^N{\cos \frac{2\pi}{n_2}k} \right) -2 \right]\\
		&\,\,  =2\varPhi \left\{ \frac{1}{2}\left[ 1+\frac{\sin \left( N+\frac{1}{2} \right) \frac{2\pi}{N}}{\sin \frac{2\pi}{2N}} \right] +\frac{1}{2}\left[ 1+\frac{\sin \left( N+\frac{1}{2} \right) \frac{2\pi}{n_2}}{\sin \frac{2\pi}{2n_2}} \right] -4 \right\}\\ 
%
		&\,\,   =2\varPhi \left[ \frac{1}{2}\left( 1+1 \right) +\frac{1}{2}\left( 1+1 \right) -4 \right] 
		\\
		&\,\,   =-4\varPhi
	\end{aligned}
\end{align}

For 3D Ising model ($N=n_1\times n_2\times n_3$),

\begin{align}\label{eq_app_4}
	\begin{aligned}	
		\sum_{p=2}^N{\varPhi \lambda _{p}^{\prime}}
		&\,\,  =\varPhi \sum_{p=2}^N{2\cos \frac{2\pi}{N}\left( p-1 \right) +2\cos \frac{2\pi n_1}{N}\left( p-1 \right)}+2\cos \frac{2\pi n_1n_2}{N}\left( p-1 \right) 
		\\
		&\,\,  =2\varPhi \sum_{k=1}^{N-1}{\cos \frac{2\pi}{N}k}+\cos \frac{2\pi}{n_2n_3}k+\cos \frac{2\pi}{n_3}k\\
		&\,\,  =2\varPhi \left[ \left( \sum_{k=0}^N{\cos \frac{2\pi}{N}k} \right) -2+\left( \sum_{k=0}^N{\cos \frac{2\pi}{n_2n_3}k} \right) -2+\left( \sum_{k=0}^N{\cos \frac{2\pi}{n_3}k} \right) -2 \right] 
		\\
		&\,\,  =2\varPhi \left\{ \frac{1}{2}\left[ 1+\frac{\sin \left( N+\frac{1}{2} \right) \frac{2\pi}{N}}{\sin \frac{2\pi}{2N}} \right] +\frac{1}{2}\left[ 1+\frac{\sin \left( N+\frac{1}{2} \right) \frac{2\pi}{n_2n_3}}{\sin \frac{2\pi}{2n_2n_3}} \right] +\frac{1}{2}\left[ 1+\frac{\sin \left( N+\frac{1}{2} \right) \frac{2\pi}{n_3}}{\sin \frac{2\pi}{2n_3}} \right] -6 \right\} 
        \\
		&\,\,   =2\varPhi \left[ \frac{1}{2}\left( 1+1 \right) +\frac{1}{2}\left( 1+1 \right) +\frac{1}{2}\left( 1+1 \right) -6 \right] 
		\\
		&\,\,   =-6\varPhi 
	\end{aligned}
\end{align}

In fact, (\ref{eq_app_2_1})-(\ref{eq_app_4}) can be obtaind in another simpler way. We illustrate this with the example of 1D model: $\sum_{p=2}^N{\varPhi \lambda _{p}^{\prime}}=-\varPhi \lambda _{1}^{\prime}+\varPhi\sum_{p=1}^N{\lambda _{p}^{\prime}}=-2\varPhi +\varPhi\ {\rm trace ({\bf A})}/z=-2\varPhi$. 

\setcounter{equation}{0} 
\section{$\frac{\log[\cosh(\lambda_p t)]}{N}$ when $\lambda_p t\rightarrow\infty$}\label{ApendE}

When $\lambda_p t=z \lambda_p^{\prime} t\rightarrow\infty$, and $\log[\cosh(z\lambda_p^{\prime} t)]\rightarrow\infty$, so $\frac{\log[\cosh(\lambda_p t)]}{N}$ in (\ref{eq12}) should be discussed. 

$I_1$ now is,

\begin{equation}\label{eq_Asymptotic_2}
	I_1=2\exp \left\{ N\left[ -\frac{1}{2}\lambda_1 t_{s1}^{2}+\log \left[\cosh \left( \lambda _1t_{s1} \right) \right] \right] \right\} \sqrt{\frac{2\pi}{\left| \lambda _1-\lambda _{1}^{2}{\rm sech}^2\left( \lambda _1t_{s1} \right) \right|}}
\end{equation}
where $t_{s1}$ is from that $t=\tanh(\lambda_1 t)$. When $\lambda_1(=z)\rightarrow \infty$, $\tanh(\lambda_1 t)\rightarrow 1$, $t_{s1}=1$, and ${\rm sech}^2(\lambda_1 t_{s1})\rightarrow 0$. Then, 

\begin{align}\label{eq_Asymptotic_3}
	\begin{aligned}	
		I_1&\rightarrow 2\exp \left\{ N\left[ -\frac{1}{2}\lambda _1+\log \left[\cosh \left( \lambda _1 \right) \right] \right] \right\} \sqrt{\frac{2\pi}{\lambda _1}}
		\\
		&=2\exp \left[ N\left( \frac{1}{2}\lambda _1-\log 2 \right) \right] \sqrt{\frac{2\pi}{\lambda _1}}
	\end{aligned}
\end{align}

For $I_p$, when $z\rightarrow\infty$, 

\begin{align}\label{eq_Asymptotic_1}
	\begin{aligned}
		I_p	&=N^{1/2}\int_{-\infty}^{+\infty} \cosh \left( \lambda _pt \right)\  {\exp \left( -\frac{1}{2}\lambda _pNt^2 \right) }\\
		&\hspace{4em}  \times \left. \exp \left\{ N\left[ \frac{2\sqrt{2}\lambda _pt}{\pi}-\log 2+\frac{2}{\pi}\int_0^{\pi /2}{\log \left( 1+e^{-2\sqrt{2}\lambda _pt\sin \xi} \right) \mathrm{d}\xi} \right] \right\} \mathrm{d}t \right.\\
		&\rightarrow N^{1/2}\int_{-\infty}^{+\infty}{\cosh \left( \lambda _pt \right) \exp \left[ N\left( -\frac{1}{2}\lambda _pt^2+\frac{2\sqrt{2}\lambda _pt}{\pi}-\log 2 \right) \right] \mathrm{d}t}\\
		&\sim 2\cosh \left(2\frac{\lambda _p}{2}{t_{sp}}\right) \exp \left[ Nf\left( \lambda _p,{t_{sp}} \right) \right]
		\sqrt{\frac{2\pi}{\lambda _p}}
	\end{aligned}
\end{align}
where $f (\lambda _p,t)=-\frac{1}{2}\lambda_pt^2+\frac{2\sqrt{2}\lambda _pt}{\pi}-\log 2$. 

When $z\rightarrow\infty$, $t_{sp}\approx 2\sqrt{2}/\pi=\bar{t}_s$. Therefore,

\begin{align}\label{eq_Asymptotic_4}
	\begin{aligned}	
		\mathcal{Z} &=\left[ \frac{\det \mathbf{K}}{\left( 2\pi \right) ^N} \right] ^{1/2} I_1\prod_{p=2}{I_p}
		\\
		&\sim 2\exp \left[ N\left( \frac{1}{2}\lambda _1 -\log 2 \right) \right] \times \prod_{p=2}{ 2\cosh \left( 2\frac{\lambda _p}{2} \bar{t}_s \right)} \exp \left[ Nf\left( \lambda _p,\bar{t}_s \right) \right]
	\end{aligned}
\end{align}

And,

\begin{align}\label{eq_Asymptotic_5}
	\begin{aligned}	
		\frac{\log \left( \mathcal{Z} \right)}{N}&=\frac{\lambda _1}{2}+\sum_{p=2}^N \left[f(\lambda_p,\bar{t}_s) \right] + \frac{1}{N}\sum_{p=2}^{N}{\log \left[ \cosh \left( 2\bar{t}_s\frac{\lambda _p}{2} \right) \right]}
	\end{aligned}
\end{align}

When $N\rightarrow\infty$ and $\lambda _p=z\lambda _{p}^{\prime}\rightarrow \infty$ but generally $z < N$, $\frac{1}{N}\sum_{p=2}^{\infty}{\log \left[ \cosh \left( 2\bar{t}_s\frac{\lambda _p}{2} \right) \right]}$ in (\ref{eq_Asymptotic_5}) is,

\begin{align}\label{eq_Asymptotic_6}
	\begin{aligned}	
		&\frac{1}{N}\sum_{p=2}^{N}{\log \left[ \cosh \left( 2\bar{t}_s\frac{\lambda _p}{2} \right) \right]}\\
		&=-\frac{(N-1)\log 2}{N}+\left( 2z\bar{t}_s \right) \frac{1}{N}\sum_{p=2}^{N}{\frac{\lambda _{p}^{\prime}}{2}}\hspace{2em}
		\\
		&=-\log 2+ \left\{ \begin{array}{ll}
			\left( 2z\bar{t}_s \right) \frac{1}{2\pi}\int_0^{2\pi}{\cos \omega _1}\mathrm{d}\omega _1,& {\rm 1D\ Model}\\
			\left( 2z\bar{t}_s \right) \frac{1}{\left( 2\pi \right) ^2}\iint\limits_0^{2\pi}{\left( \cos \omega _1+\cos \omega _2 \right) \mathrm{d}\omega _1\mathrm{d}\omega _2},& {\rm 2D\ Model}\\
			\left( 2z\bar{t}_s \right) \frac{1}{\left( 2\pi \right) ^3}\iiint\limits_0^{2\pi}{\left( \cos \omega _1+\cos \omega _2+\cos \omega _3 \right)}\mathrm{d}\omega _1\mathrm{d}\omega _2\mathrm{d}\omega _3,& {\rm 3D\ Model}
		\end{array} \right.
		\\
		&=-\log 2
	\end{aligned}
\end{align}

Therefore in the case that $\lambda_p t=z \lambda_p^{\prime} t\rightarrow\infty$, $\frac{\log[\cosh(\lambda_p t)]}{N}$ in (\ref{eq12}) can also be omitted, because only one constant "$-\log 2$" will be added to the final result \footnote{\tiny If $z$ and $N$ tend to infinity at the same time (ie., $z/N\rightarrow 1$), the constants are $-\log 2 -2\bar{t}_s$, $-\log 2 -4\bar{t}_s$ and $-\log 2 -6\bar{t}_s$ for 1D-3D Ising model, respectively.}.

\setcounter{equation}{0} 
\section{Approximating $\mathcal{I}\left( \lambda _p,t \right)$ by economization of power series}\label{Apend_preD}

From  Taylor expansion of $\log \left[ \cosh \left( \sqrt{2}\lambda _pt\sin \xi \right) \right](=\log \left[\cosh \left(x_0\bar{\lambda}_p t^\prime \sin \xi\right)\right])$  in the neighborhood of $x_0$ ($=\eta \sqrt{2}z$), we can approximate $\mathcal{I}\left( \lambda _p,t \right)$. That is,

\begin{align}\label{SCeq4}
	\begin{aligned}
		\mathcal{I}\left( \lambda _p,t \right)&=\frac{2}{\pi}\int_0^{\pi /2}{\log \left\{ \cosh \left[ x_0\bar{\lambda}_pt^{\prime}\sin \xi \right] \right\} \mathrm{d}\xi}\\
		&\approx \frac{2}{\pi}\int_0^{\pi /2}{\left[ A_0(x_0)+A_1(x_0)\left( x_0\bar{\lambda}_pt^{\prime}\sin \xi -x_0 \right) +...+A_N(x_0)\left( x_0\bar{\lambda}_pt^{\prime}\sin \xi -x_0 \right)^N \right]\mathrm{d}\xi}\\
	\end{aligned}
\end{align}
where $A_{k}(x_0)$ is the $k$-th coefficient of Taylor series of $\log[\cosh(x)]$ at $x_0$, such as that $A_0(x_0)=\log[\cosh (x_0)]$, that $A_1(x_0)=\tanh (x_0)$, that $A_2(x_0)=\frac{1}{2}{\rm sech}^2(x_0)$, that $A_3(x_0)=\frac{1}{3!}[-2{\rm sech}^2(x_0)\tanh(x_0)]$, that $A_4(x_0)=\frac{1}{4!}[-2{\rm sech}^4(x_0)+4{\rm sech}^2(x_0)\tanh^2(x_0)]$, and so on. Clearly, a good approximation of $\log[\cosh(x)]$ requires theoretically a large number of terms of its Taylor series (e.g., $N=28$) if the numerical stability is not taken into account. Note that this series is assumed to be convergent for simplicity.

However, we can approximate $\log[\cosh(x_0\bar{\lambda}_p t^\prime \sin \xi)]$ by a 2-order series instead of a 28-order Taylor series with some acceptable error using the technique of economization of power series. That is,  

\begin{align}\label{SCeq4_1}
	\begin{aligned}
		&\frac{2}{\pi}\int_0^{\pi /2}{\log \left\{ \cosh \left[ x_0\bar{\lambda}_pt^{\prime}\sin \xi \right] \right\} \mathrm{d}\xi}\\
		&\approx \frac{2}{\pi}\int_0^{\pi /2}{\left[ F_0(x_0)+F_1(x_0)\left( x_0\bar{\lambda}_pt^{\prime}\sin \xi -x_0 \right) +F_2(x_0)\left( x_0\bar{\lambda}_pt^{\prime}\sin \xi -x_0 \right) ^2 \right]\mathrm{d}\xi}\\
	\end{aligned}
\end{align}
where $F_{k}(x_0)(k=0,1,2)$ are calculated step by step as follows,

\begin{equation}\label{SCeq4_2}
	\left\{ \begin{array}{l}
		F_2(x_0)=E_2(x_0)\\
		F_1(x_0)=E_1(x_0)+2F_2(x_0)x_0\\
		F_0(x_0)=E_0(x_0)+F_1(x_0)-F_2(x_0)x_0^2\\
	\end{array} \right. 	
\end{equation}

\begin{equation}\label{SCeq4_2_1}
	\left\{ \begin{array}{l}
		E_2(x_0)=8D_2(x_0)\\
		E_1(x_0)=2D_1(x_0)-8D_2(x_0)\\
		E_0(x_0)=D_0(x_0)-D_1(x_0)+D_2(x_0)\\
	\end{array} \right. 	
\end{equation}

We abbreviate $D_2(x_0)$ to $D_2$ (The same to others) in the following,  

\begin{equation}\label{SCeq4_3}
	\left\{ \begin{array}{l}
		D_0=C_0+\frac{1}{2}C_2+\frac{3}{8}C_4+\frac{5}{16}C_6+\frac{35}{128}C_8+\frac{63}{256}C_{10}+\frac{231}{1024}C_{12}+\frac{429}{2048}C_{14}+\frac{6435}{32768}C_{16}+\frac{12155}{65536}C_{18}\\
		\hspace{3em}+\frac{46189}{262144}C_{20}+\frac{88179}{524288}C_{22}+\frac{676039}{4194304}C_{24}+\frac{1300075}{8388608}C_{26}+\frac{5014575}{33554432}C_{28}\\
		D_1=C_1+\frac{3}{4}C_3+\frac{5}{8}C_5+\frac{35}{64}C_7+\frac{63}{128}C_9+\frac{231}{512}C_{11}+\frac{429}{1024}C_{13}+\frac{6435}{16384}C_{15}+\frac{12155}{32768}C_{17}+\frac{46189}{131072}C_{19}\\
		\hspace{3em}+\frac{88179}{262144}C_{21}+\frac{676039}{2097152}C_{23}+\frac{1300075}{4194304}C_{25}+\frac{5014575}{16777216}C_{27}\\		
		D_2=\frac{1}{2}C_2+\frac{1}{2}C_4+\frac{15}{32}C_6+\frac{7}{16}C_8+\frac{105}{256}C_{10}+\frac{99}{256}C_{12}+\frac{3003}{8192}C_{14}+\frac{715}{2048}C_{16}+\frac{21879}{65536}C_{18}+\frac{20995}{65536}C_{20}\\
		\hspace{3em}+\frac{323323}{1048576}C_{22}+\frac{156009}{524288}C_{24}+\frac{2414425}{8388608}C_{26}+\frac{2340135}{8388608}C_{28}		
	\end{array} \right.
\end{equation}

\begin{align}\label{SCeq4_4}
	\begin{aligned}
		\left\{ \begin{array}{l}
			C_0=B_0+\frac{B_1}{2}+\frac{B_2}{4}+\frac{B_3}{8}+\frac{B_4}{16}+\frac{B_5}{32}+\frac{B_6}{64}+\frac{B_7}{128}+\frac{B_8}{256}+\frac{B_9}{512}+\frac{B_{10}}{1024}+\frac{B_{11}}{2048}+\frac{B_{12}}{4096}+\frac{B_{13}}{8192}+\frac{B_{14}}{16384}\\
			\hspace{2em}+\frac{B_{15}}{32768}+\frac{B_{16}}{65536}+\frac{B_{17}}{131072}+\frac{B_{18}}{262144}+\frac{B_{19}}{524288}+\frac{B_{20}}{1048576}+\frac{B_{21}}{2097152}+\frac{B_{22}}{4194304}+\frac{B_{23}}{8388608}+\frac{B_{24}}{16777216}\\
			\hspace{2em}+\frac{B_{25}}{33554432}+\frac{B_{26}}{67108864}+\frac{B_{27}}{134217728}+\frac{B_{28}}{268435456}			
			\\
			C_1=\frac{B_1}{2}+\frac{B_2}{2}+\frac{3B_3}{8}+\frac{B_4}{4}+\frac{5B_5}{32}+\frac{3B_6}{32}+\frac{7B_7}{128}+\frac{B_8}{32}+\frac{9B_9}{512}+\frac{5B_{10}}{512}+\frac{11B_{11}}{2048}+\frac{3B_{12}}{1024}+\frac{13B_{13}}{8192}+\frac{7B_{14}}{8192}\\
			\hspace{2em}+\frac{15B_{15}}{32768}+\frac{B_{16}}{4096}+\frac{17B_{17}}{131072}+\frac{9B_{18}}{131072}+\frac{19B_{19}}{524288}+\frac{5B_{20}}{262144}+\frac{21B_{21}}{2097152}+\frac{11B_{22}}{2097152}+\frac{23B_{23}}{8388608}+\frac{3B_{24}}{2097152}\\
			\hspace{2em}+\frac{25B_{25}}{33554432}+\frac{13B_{26}}{33554432}+\frac{27B_{27}}{134217728}+\frac{7B_{28}}{67108864}			
			\\
			C_2=\frac{B_2}{4}+\frac{3B_3}{8}+\frac{3B_4}{8}+\frac{5B_5}{16}+\frac{15B_6}{64}+\frac{21B_7}{128}+\frac{7B_8}{64}+\frac{9B_9}{128}+\frac{45B_{10}}{1024}+\frac{55B_{11}}{2048}+\frac{33B_{12}}{2048}+\frac{39B_{13}}{4096}+\frac{91B_{14}}{16384}\\
			\hspace{2em}+\frac{105B_{15}}{32768}+\frac{15B_{16}}{8192}+\frac{17B_{17}}{16384}+\frac{153B_{18}}{262144}+\frac{171B_{19}}{524288}+\frac{95B_{20}}{524288}+\frac{105B_{21}}{1048576}+\frac{231B_{22}}{4194304}+\frac{253B_{23}}{8388608}+\frac{69B_{24}}{4194304}\\
			\hspace{2em}+\frac{75B_{25}}{8388608}+\frac{325B_{26}}{67108864}+\frac{351B_{27}}{134217728}+\frac{189B_{28}}{134217728}			
			\\
			C_3=\frac{B_3}{8}+\frac{B_4}{4}+\frac{5B_5}{16}+\frac{5B_6}{16}+\frac{35B_7}{128}+\frac{7B_8}{32}+\frac{21B_9}{128}+\frac{15B_{10}}{128}+\frac{165B_{11}}{2048}+\frac{55B_{12}}{1024}+\frac{143B_{13}}{4096}+\frac{91B_{14}}{4096}\\
			\hspace{2em}+\frac{455B_{15}}{32768}+\frac{35B_{16}}{4096}+\frac{85B_{17}}{16384}+\frac{51B_{18}}{16384}+\frac{969B_{19}}{524288}+\frac{285B_{20}}{262144}+\frac{665B_{21}}{1048576}+\frac{385B_{22}}{1048576}+\frac{1771B_{23}}{8388608}+\frac{253B_{24}}{2097152}\\
			\hspace{2em}+\frac{575B_{25}}{8388608}+\frac{325B_{26}}{8388608}+\frac{2925B_{27}}{134217728}+\frac{819B_{28}}{67108864}
		 \end{array} \right.
		\end{aligned}
	\end{align}
	
	\setcounter{equation}{5}
	
	\begin{align}\label{SCeq4_4}
		\begin{aligned}
			\left\{ \begin{array}{l}			
			\\
			C_4=\frac{B_4}{16}+\frac{5B_5}{32}+\frac{15B_6}{64}+\frac{35B_7}{128}+\frac{35B_8}{128}+\frac{63B_9}{256}+\frac{105B_{10}}{512}+\frac{165B_{11}}{1024}+\frac{495B_{12}}{4096}+\frac{715B_{13}}{8192}+\frac{1001B_{14}}{16384}\\
			\hspace{2em}+\frac{1365B_{15}}{32768}+\frac{455B_{16}}{16384}+\frac{595B_{17}}{32768}+\frac{765B_{18}}{65536}+\frac{969B_{19}}{131072}+\frac{4845B_{20}}{1048576}+\frac{5985B_{21}}{2097152}+\frac{7315B_{22}}{4194304}+\frac{8855B_{23}}{8388608}\\
			\hspace{2em}+\frac{5313B_{24}}{8388608}+\frac{6325B_{25}}{16777216}+\frac{7475B_{26}}{33554432}+\frac{8775B_{27}}{67108864}+\frac{20475B_{28}}{268435456}
			\\
			C_5=\frac{B_5}{32}+\frac{3B_6}{32}+\frac{21B_7}{128}+\frac{7B_8}{32}+\frac{63B_9}{256}+\frac{63B_{10}}{256}+\frac{231B_{11}}{1024}+\frac{99B_{12}}{512}+\frac{1287B_{13}}{8192}+\frac{1001B_{14}}{8192}+\frac{3003B_{15}}{32768}\\
			\hspace{2em}+\frac{273B_{16}}{4096}+\frac{1547B_{17}}{32768}+\frac{1071B_{18}}{32768}+\frac{2907B_{19}}{131072}+\frac{969B_{20}}{65536}+\frac{20349B_{21}}{2097152}+\frac{13167B_{22}}{2097152}+\frac{33649B_{23}}{8388608}\\
			\hspace{2em}+\frac{5313B_{24}}{2097152}+\frac{26565B_{25}}{16777216}+\frac{16445B_{26}}{16777216}+\frac{40365B_{27}}{67108864}+\frac{12285B_{28}}{33554432}\\
			C_6=\frac{B_6}{64}+\frac{7B_7}{128}+\frac{7B_8}{64}+\frac{21B_9}{128}+\frac{105B_{10}}{512}+\frac{231B_{11}}{1024}+\frac{231B_{12}}{1024}+\frac{429B_{13}}{2048}+\frac{3003B_{14}}{16384}+\frac{5005B_{15}}{32768}\\
			\hspace{2em}+\frac{1001B_{16}}{8192}+\frac{1547B_{17}}{16384}+\frac{4641B_{18}}{65536}+\frac{6783B_{19}}{131072}+\frac{4845B_{20}}{131072}+\frac{6783B_{21}}{262144}+\frac{74613B_{22}}{4194304}+\frac{100947B_{23}}{8388608}\\
			\hspace{2em}+\frac{33649B_{24}}{4194304}+\frac{44275B_{25}}{8388608}+\frac{115115B_{26}}{33554432}+\frac{148005B_{27}}{67108864}+\frac{94185B_{28}}{67108864}
			\\
			C_7=\frac{B_7}{128}+\frac{B_8}{32}+\frac{9B_9}{128}+\frac{15B_{10}}{128}+\frac{165B_{11}}{1024}+\frac{99B_{12}}{512}+\frac{429B_{13}}{2048}+\frac{429B_{14}}{2048}+\frac{6435B_{15}}{32768}+\frac{715B_{16}}{4096}\\
			\hspace{2em}+\frac{2431B_{17}}{16384}+\frac{1989B_{18}}{16384}+\frac{12597B_{19}}{131072}+\frac{4845B_{20}}{65536}+\frac{14535B_{21}}{262144}+\frac{10659B_{22}}{262144}+\frac{245157B_{23}}{8388608}+\frac{43263B_{24}}{2097152}\\
			\hspace{2em}+\frac{120175B_{25}}{8388608}+\frac{82225B_{26}}{8388608}+\frac{444015B_{27}}{67108864}+\frac{148005B_{28}}{33554432}			
		\end{array} \right.
	\end{aligned}
\end{align}

\setcounter{equation}{5}

\begin{align*}
	\begin{aligned}
		\left\{ \begin{array}{l}
			C_8=\frac{B_8}{256}+\frac{9B_9}{512}+\frac{45B_{10}}{1024}+\frac{165B_{11}}{2048}+\frac{495B_{12}}{4096}+\frac{1287B_{13}}{8192}+\frac{3003B_{14}}{16384}+\frac{6435B_{15}}{32768}+\frac{6435B_{16}}{32768}+\frac{12155B_{17}}{65536}\\
			\hspace{2em}+\frac{21879B_{18}}{131072}+\frac{37791B_{19}}{262144}+\frac{62985B_{20}}{524288}+\frac{101745B_{21}}{1048576}+\frac{159885B_{22}}{2097152}+\frac{245157B_{23}}{4194304}+\frac{735471B_{24}}{16777216}\\
			\hspace{2em}+\frac{1081575B_{25}}{33554432}+\frac{1562275B_{26}}{67108864}+\frac{2220075B_{27}}{134217728}+\frac{3108105B_{28}}{268435456}			
			\\
			C_9=\frac{B_9}{512}+\frac{5B_{10}}{512}+\frac{55B_{11}}{2048}+\frac{55B_{12}}{1024}+\frac{715B_{13}}{8192}+\frac{1001B_{14}}{8192}+\frac{5005B_{15}}{32768}+\frac{715B_{16}}{4096}+\frac{12155B_{17}}{65536}+\frac{12155B_{18}}{65536}\\
			\hspace{2em}+\frac{46189B_{19}}{262144}+\frac{20995B_{20}}{131072}+\frac{146965B_{21}}{1048576}+\frac{124355B_{22}}{1048576}+\frac{408595B_{23}}{4194304}+\frac{81719B_{24}}{1048576}+\frac{2042975B_{25}}{33554432}
			\\
			\hspace{2em}+\frac{1562275B_{26}}{33554432}+\frac{4686825B_{27}}{134217728}+\frac{1726725B_{28}}{67108864}			
			\\
			C_{10}=\frac{B_{10}}{1024}+\frac{11B_{11}}{2048}+\frac{33B_{12}}{2048}+\frac{143B_{13}}{4096}+\frac{1001B_{14}}{16384}+\frac{3003B_{15}}{32768}+\frac{1001B_{16}}{8192}+\frac{2431B_{17}}{16384}+\frac{21879B_{18}}{131072}\\
			\hspace{2em}+\frac{46189B_{19}}{262144}+\frac{46189B_{20}}{262144}+\frac{88179B_{21}}{524288}+\frac{323323B_{22}}{2097152}+\frac{572033B_{23}}{4194304}+\frac{245157B_{24}}{2097152}
			\\
			\hspace{2em}+\frac{408595B_{25}}{4194304}+\frac{5311735B_{26}}{67108864}+\frac{8436285B_{27}}{134217728}+\frac{6561555B_{28}}{134217728}\\				C_{11}=\frac{B_{11}}{2048}+\frac{3B_{12}}{1024}+\frac{39B_{13}}{4096}+\frac{91B_{14}}{4096}+\frac{1365B_{15}}{32768}+\frac{273B_{16}}{4096}+\frac{1547B_{17}}{16384}+\frac{1989B_{18}}{16384}\\
			\hspace{2em}+\frac{37791B_{19}}{262144}+\frac{20995B_{20}}{131072}+\frac{88179B_{21}}{524288}+\frac{88179B_{22}}{524288}+\frac{676039B_{23}}{4194304}+\frac{156009B_{24}}{1048576}
			\\
			\hspace{2em}+\frac{557175B_{25}}{4194304}+\frac{482885B_{26}}{4194304}+\frac{13037895B_{27}}{134217728}+\frac{5368545B_{28}}{67108864}			
		\end{array} \right.
	\end{aligned}
\end{align*}
\setcounter{equation}{5}

\begin{align*}
	\begin{aligned}
		\left\{ \begin{array}{l}
			C_{12}=\frac{B_{12}}{4096}+\frac{13B_{13}}{8192}+\frac{91B_{14}}{16384}+\frac{455B_{15}}{32768}+\frac{455B_{16}}{16384}+\frac{1547B_{17}}{32768}+\frac{4641B_{18}}{65536}+\frac{12597B_{19}}{131072}+\frac{62985B_{20}}{524288}+\frac{146965B_{21}}{1048576}\\
			\hspace{2em}+\frac{323323B_{22}}{2097152}+\frac{676039B_{23}}{4194304}+\frac{676039B_{24}}{4194304}
			+\frac{1300075B_{25}}{8388608}+\frac{2414425B_{26}}{16777216}+\frac{4345965B_{27}}{33554432}+\frac{30421755B_{28}}{268435456}			
			\\
			C_{13}=\frac{B_{13}}{8192}+\frac{7B_{14}}{8192}+\frac{105B_{15}}{32768}+\frac{35B_{16}}{4096}+\frac{595B_{17}}{32768}+\frac{1071B_{18}}{32768}+\frac{6783B_{19}}{131072}+\frac{4845B_{20}}{65536}+\frac{101745B_{21}}{1048576}+\frac{124355B_{22}}{1048576}\\
			\hspace{2em}+\frac{572033B_{23}}{4194304}+\frac{156009B_{24}}{1048576}+\frac{1300075B_{25}}{8388608}+\frac{1300075B_{26}}{8388608}+\frac{5014575B_{27}}{33554432}+\frac{2340135B_{28}}{16777216}			
			\\
			C_{14}=\frac{B_{14}}{16384}+\frac{15B_{15}}{32768}+\frac{15B_{16}}{8192}+\frac{85B_{17}}{16384}+\frac{765B_{18}}{65536}+\frac{2907B_{19}}{131072}+\frac{4845B_{20}}{131072}+\frac{14535B_{21}}{262144}+\frac{159885B_{22}}{2097152}+\frac{408595B_{23}}{4194304}\\
			\hspace{2em}+\frac{245157B_{24}}{2097152}+\frac{557175B_{25}}{4194304}+\frac{2414425B_{26}}{16777216}+\frac{5014575B_{27}}{33554432}+\frac{5014575B_{28}}{33554432}
			\\
			C_{15}=\frac{B_{15}}{32768}+\frac{B_{16}}{4096}+\frac{17B_{17}}{16384}+\frac{51B_{18}}{16384}+\frac{969B_{19}}{131072}+\frac{969B_{20}}{65536}+\frac{6783B_{21}}{262144}+\frac{10659B_{22}}{262144}+\frac{245157B_{23}}{4194304}+\frac{81719B_{24}}{1048576}\\
			\hspace{2em}+\frac{408595B_{25}}{4194304}+\frac{482885B_{26}}{4194304}+\frac{4345965B_{27}}{33554432}+\frac{2340135B_{28}}{16777216}
				\end{array} \right.
		\end{aligned}
	\end{align*}
\setcounter{equation}{5}
\begin{align}
	\begin{aligned}
		\left\{ \begin{array}{l}
			C_{16}=\frac{B_{16}}{65536}+\frac{17B_{17}}{131072}+\frac{153B_{18}}{262144}+\frac{969B_{19}}{524288}+\frac{4845B_{20}}{1048576}+\frac{20349B_{21}}{2097152}+\frac{74613B_{22}}{4194304}+\frac{245157B_{23}}{8388608}+\frac{735471B_{24}}{16777216}\\
			\hspace{2em}+\frac{2042975B_{25}}{33554432}+\frac{5311735B_{26}}{67108864}+\frac{13037895B_{27}}{134217728}+\frac{30421755B_{28}}{268435456}			
			\\
			C_{17}=\frac{B_{17}}{131072}+\frac{9B_{18}}{131072}+\frac{171B_{19}}{524288}+\frac{285B_{20}}{262144}+\frac{5985B_{21}}{2097152}+\frac{13167B_{22}}{2097152}+\frac{100947B_{23}}{8388608}+\frac{43263B_{24}}{2097152}+\frac{1081575B_{25}}{33554432}\\
			\hspace{2em}+\frac{1562275B_{26}}{33554432}+\frac{8436285B_{27}}{134217728}+\frac{5368545B_{28}}{67108864}			
			\\
			C_{18}=\frac{B_{18}}{262144}+\frac{19B_{19}}{524288}+\frac{95B_{20}}{524288}+\frac{665B_{21}}{1048576}+\frac{7315B_{22}}{4194304}+\frac{33649B_{23}}{8388608}+\frac{33649B_{24}}{4194304}+\frac{120175B_{25}}{8388608}+\frac{1562275B_{26}}{67108864}\\
			\hspace{2em}+\frac{4686825B_{27}}{134217728}+\frac{6561555B_{28}}{134217728}
			\\
			C_{19}=\frac{B_{19}}{524288}+\frac{5B_{20}}{262144}+\frac{105B_{21}}{1048576}+\frac{385B_{22}}{1048576}+\frac{8855B_{23}}{8388608}+\frac{5313B_{24}}{2097152}+\frac{44275B_{25}}{8388608}+\frac{82225B_{26}}{8388608}+\frac{2220075B_{27}}{134217728}\\
			\hspace{2em}+\frac{1726725B_{28}}{67108864}
				\end{array} \right.
		\end{aligned}
	\end{align}
\setcounter{equation}{5}
      \begin{align*}
	    \begin{aligned}
		\left\{ \begin{array}{l}
			C_{20}=\frac{B_{20}}{1048576}+\frac{21B_{21}}{2097152}+\frac{231B_{22}}{4194304}+\frac{1771B_{23}}{8388608}+\frac{5313B_{24}}{8388608}+\frac{26565B_{25}}{16777216}+\frac{115115B_{26}}{33554432}+\frac{444015B_{27}}{67108864}+\frac{3108105B_{28}}{268435456}\\
			C_{21}=\frac{B_{21}}{2097152}+\frac{11B_{22}}{2097152}+\frac{253B_{23}}{8388608}+\frac{253B_{24}}{2097152}+\frac{6325B_{25}}{16777216}+\frac{16445B_{26}}{16777216}+\frac{148005B_{27}}{67108864}+\frac{148005B_{28}}{33554432}\\
			C_{22}=\frac{B_{22}}{4194304}+\frac{23B_{23}}{8388608}+\frac{69B_{24}}{4194304}+\frac{575B_{25}}{8388608}+\frac{7475B_{26}}{33554432}+\frac{40365B_{27}}{67108864}+\frac{94185B_{28}}{67108864}\\
			C_{23}=\frac{B_{23}}{8388608}+\frac{3B_{24}}{2097152}+\frac{75B_{25}}{8388608}+\frac{325B_{26}}{8388608}+\frac{8775B_{27}}{67108864}+\frac{12285B_{28}}{33554432}\\
			C_{24}=\frac{B_{24}}{16777216}+\frac{25B_{25}}{33554432}+\frac{325B_{26}}{67108864}+\frac{2925B_{27}}{134217728}+\frac{20475B_{28}}{268435456}\\
			C_{25}=\frac{B_{25}}{33554432}+\frac{13B_{26}}{33554432}+\frac{351B_{27}}{134217728}+\frac{819B_{28}}{67108864}\\
			C_{26}=\frac{B_{26}}{67108864}+\frac{27B_{27}}{134217728}+\frac{189B_{28}}{134217728}\\
			C_{27}=\frac{B_{27}}{134217728}+\frac{7B_{28}}{67108864}\\
			C_{28}=\frac{B_{28}}{268435456}					
		\end{array} \right.
	\end{aligned}
\end{align*}

In the following $A_{k}$ is the $k$-th coefficient of Taylor series of $\log[\cosh(x)]$ at $x_0$ (See (\ref{SCeq4})).

\setcounter{equation}{6}

\begin{align}\label{SCeq4_5}
	\begin{aligned}
		\left\{ \begin{array}{l}
			B_0=A_0-A_1x_0+A_2x_{0}^{2}-A_3x_{0}^{3}+A_4x_{0}^{4}-A_5x_{0}^{5}+A_6x_{0}^{6}-A_7x_{0}^{7}+A_8x_{0}^{8}-A_9x_{0}^{9}+A_{10}x_{0}^{10}-A_{11}x_{0}^{11}
			\\
			\hspace{2.5em}+A_{12}x_{0}^{12}-A_{13}x_{0}^{13}+A_{14}x_{0}^{14}-A_{15}x_{0}^{15}+A_{16}x_{0}^{16}-A_{17}x_{0}^{17}+A_{18}x_{0}^{18}-A_{19}x_{0}^{19}+A_{20}x_{0}^{20}-A_{21}x_{0}^{21}
			\\
			\hspace{2.5em}+A_{22}x_{0}^{22}-A_{23}x_{0}^{23}+A_{24}x_{0}^{24}-A_{25}x_{0}^{25}+A_{26}x_{0}^{26}-A_{27}x_{0}^{27}+A_{28}x_{0}^{28}			
			\\
			B_1=A_1-2A_2x_0+3A_3x_{0}^{2}-4A_4x_{0}^{3}+5A_5x_{0}^{4}-6A_6x_{0}^{5}+7A_7x_{0}^{6}-8A_8x_{0}^{7}+9A_9x_{0}^{8}-10A_{10}x_{0}^{9}
			\\
			\hspace{2.5em}+11A_{11}x_{0}^{10}-12A_{12}x_{0}^{11}+13A_{13}x_{0}^{12}-14A_{14}x_{0}^{13}+15A_{15}x_{0}^{14}-16A_{16}x_{0}^{15}+17A_{17}x_{0}^{16}-18A_{18}x_{0}^{17}
			\\
			\hspace{2.5em}+19A_{19}x_{0}^{18}-20A_{20}x_{0}^{19}+21A_{21}x_{0}^{20}-22A_{22}x_{0}^{21}+23A_{23}x_{0}^{22}-24A_{24}x_{0}^{23}+25A_{25}x_{0}^{24}-26A_{26}x_{0}^{25}\\
			\hspace{2.5em}+27A_{27}x_{0}^{26}-28A_{28}x_{0}^{27}
			\\
			B_2=A_2-3A_3x_0+6A_4x_{0}^{2}-10A_5x_{0}^{3}+15A_6x_{0}^{4}-21A_7x_{0}^{5}+28A_8x_{0}^{6}-36A_9x_{0}^{7}+45A_{10}x_{0}^{8}-55A_{11}x_{0}^{9}
			\\
			\hspace{2.5em}+66A_{12}x_{0}^{10}-78A_{13}x_{0}^{11}+91A_{14}x_{0}^{12}-105A_{15}x_{0}^{13}+120A_{16}x_{0}^{14}-136A_{17}x_{0}^{15}+153A_{18}8x_{0}^{16}
			\\
			\hspace{2.5em}-171A_{19}x_{0}^{17}+190A_{20}x_{0}^{18}-210A_{21}x_{0}^{19}+231A_{22}x_{0}^{20}-253A_{23}x_{0}^{21}+276A_{24}x_{0}^{22}-300A_{25}x_{0}^{23}\\
			\hspace{2.5em}+325A_{26}x_{0}^{24}-351A_{27}x_{0}^{25}+378A_{28}x_{0}^{26}
			\\
			B_3=A_3-4A_4x_0+10A_5x_{0}^{2}-20A_6x_{0}^{3}+35A_7x_{0}^{4}-56A_8x_{0}^{5}+84A_9x_{0}^{6}-120A_{10}x_{0}^{7}+165A_{11}x_{0}^{8}
			\\
			\hspace{2.5em}-220A_{12}x_{0}^{9}+286A_{13}x_{0}^{10}-364A_{14}x_{0}^{11}+455A_{15}x_{0}^{12}-560A_{16}x_{0}^{13}+680A_{17}x_{0}^{14}-816A_{18}x_{0}^{15}
			\\
			\hspace{2.5em}+969A_{19}x_{0}^{16}-1140A_{20}x_{0}^{17}+1330A_{21}x_{0}^{18}-1540A_{22}x_{0}^{19}+1771A_{23}x_{0}^{20}-2024A_{24}x_{0}^{21}
			\\
			\hspace{2.5em}+2300A_{25}x_{0}^{22}-2600A_{26}x_{0}^{23}+2925A_{27}x_{0}^{24}-3276A_{28}x_{0}^{25}
		\end{array} \right.
		\end{aligned}
	\end{align}
\setcounter{equation}{6}
	\begin{align}\label{SCeq4_5}
			\begin{aligned} 
			 \left\{ \begin{array}{l}
			B_4=A_4-5A_5x_0+15A_6x_{0}^{2}-35A_7x_{0}^{3}+70A_8x_{0}^{4}-126A_9x_{0}^{5}+210A_10x_{0}^{6}-330A_{11}x_{0}^{7}+495A_{12}x_{0}^{8}
			\\
			\hspace{2.5em}-715A_{13}x_{0}^{9}+1001A_{14}x_{0}^{10}-1365A_{15}x_{0}^{11}+1820A_{16}x_{0}^{12}-2380A_{17}x_{0}^{13}+3060A_{18}x_{0}^{14}
			\\
			\hspace{2.5em}-3876A_{19}x_{0}^{15}+4845A_{20}x_{0}^{16}-5985A_{21}x_{0}^{17}+7315A_{22}x_{0}^{18}-8855A_{23}x_{0}^{19}+10626A_{24}x_{0}^{20}
			\\
			\hspace{2.5em}-12650A_{25}x_{0}^{21}+14950A_{26}x_{0}^{22}-17550A_{27}x_{0}^{23}+20475A_{28}x_{0}^{24}			
			\\
			B_5=A_5-6A_6x_0+21A_7x_{0}^{2}-56A_8x_{0}^{3}+126A_9x_{0}^{4}-252A_{10}x_{0}^{5}+462A_{11}x_{0}^{6}-792A_{12}x_{0}^{7}+1287A_{13}x_{0}^{8}
			\\
			\hspace{2.5em}-2002A_{14}x_{0}^{9}+3003A_{15}x_{0}^{10}-4368A_{16}x_{0}^{11}+6188A_{17}x_{0}^{12}-8568A_{18}x_{0}^{13}+11628A_{19}x_{0}^{14}
			\\
			\hspace{2.5em}-15504A_{20}x_{0}^{15}+20349A_{21}x_{0}^{16}-26334A_{22}x_{0}^{17}+33649A_{23}x_{0}^{18}-42504A_{24}x_{0}^{19}
			\\
			\hspace{2.5em}+53130A_{25}x_{0}^{20}-65780A_{26}x_{0}^{21}+80730A_{27}x_{0}^{22}-98280A_{28}x_{0}^{23}
			\\
			B_6=A_6-7A_7x_0+28A_8x_{0}^{2}-84A_9x_{0}^{3}+210A_{10}x_{0}^{4}-462A_{11}x_{0}^{5}+924A_{12}x_{0}^{6}-1716A_{13}x_{0}^{7}
			\\
			\hspace{2.5em}+3003A_{14}x_{0}^{8}-5005A_{15}x_{0}^{9}+8008A_{16}x_{0}^{10}-12376A_{17}x_{0}^{11}+18564A_{18}x_{0}^{12}-27132A_{19}x_{0}^{13}
			\\
			\hspace{2.5em}+38760A_{20}x_{0}^{14}-54264A_{21}x_{0}^{15}+74613A_{22}x_{0}^{16}-100947A_{23}x_{0}^{17}+134596A_{24}x_{0}^{18}
			\\
			\hspace{2.5em}-177100A_{25}x_{0}^{19}+230230A_{26}x_{0}^{20}-296010A_{27}x_{0}^{21}+376740A_{28}x_{0}^{22}			
			\\
			B_7=A_7-8A_8x_0+36A_9x_{0}^{2}-120A_{10}x_{0}^{3}+330A_{11}x_{0}^{4}-792A_{12}x_{0}^{5}+1716A_{13}x_{0}^{6}-3432A_{14}x_{0}^{7}
			\\
			\hspace{2.5em}+6435A_{15}x_{0}^{8}-11440A_{16}x_{0}^{9}+19448A_{17}x_{0}^{10}-31824A_{18}x_{0}^{11}+50388A_{19}x_{0}^{12}-77520A_{20}x_{0}^{13}
			\\
			\hspace{2.5em}+116280A_{21}x_{0}^{14}-170544A_{22}x_{0}^{15}+245157A_{23}x_{0}^{16}-346104A_{24}x_{0}^{17}+480700A_{25}x_{0}^{18}
			\\
			\hspace{2.5em}-657800A_{26}x_{0}^{19}+888030A_{27}x_{0}^{20}-1184040A_{28}x_{0}^{21}			
		 \end{array} \right.
		\end{aligned}
	\end{align}

\setcounter{equation}{6}
\begin{align*}
	\begin{aligned}
		\left\{ \begin{array}{l}
			B_8=A_8-9A_9x_0+45A_{10}x_{0}^{2}-165A_{11}x_{0}^{3}+495A_{12}x_{0}^{4}-1287A_{13}x_{0}^{5}+3003A_{14}x_{0}^{6}-6435A_{15}x_{0}^{7}
			\\
			\hspace{2.5em}+12870A_{16}x_{0}^{8}-24310A_{17}x_{0}^{9}+43758A_{18}x_{0}^{1}0-75582A_{19}x_{0}^{11}+125970A_{20}x_{0}^{12}
			\\
			\hspace{2.5em}-203490A_{21}x_{0}^{13}+319770A_{22}x_{0}^{14}-490314A_{23}x_{0}^{15}+735471A_{24}x_{0}^{16}
			\\
			\hspace{2.5em}-1081575A_{25}x_{0}^{17}+1562275A_{26}x_{0}^{18}-2220075A_{27}x_{0}^{19}+3108105A_{28}x_{0}^{20}
			\\
			B_9=A_9-10A_{10}x_0+55A_{11}x_{0}^{2}-220A_{12}x_{0}^{3}+715A_{13}x_{0}^{4}-2002A_{14}x_{0}^{5}+5005A_{15}x_{0}^{6}-11440A_{16}x_{0}^{7}
			\\
			\hspace{2.5em}+24310A_{17}x_{0}^{8}-48620A_{18}x_{0}^{9}+92378A_{19}x_{0}^{10}-167960A_{20}x_{0}^{11}+293930A_{21}x_{0}^{12}
			\\
			\hspace{2.5em}-497420A_{23}x_{0}^{13}+817190A_{23}x_{0}^{14}-1307504A_{24}x_{0}^{15}+2042975A_{25}x_{0}^{16}
			\\
			\hspace{2.5em}-3124550A_{26}x_{0}^{17}+4686825A_{27}x_{0}^{18}-6906900A_{28}x_{0}^{19}
			\\
			B_{10}=A_{10}-11A_{11}x_0+66A_{12}x_{0}^{2}-286A_{13}x_{0}^{3}+1001A_{14}x_{0}^{4}-3003A_{15}x_{0}^{5}+8008A_{16}x_{0}^{6}
			\\
			\hspace{2.5em}-19448A_{17}x_{0}^{7}+43758A_{18}x_{0}^{8}-92378A_{19}x_{0}^{9}+184756A_{20}x_{0}^{10}-352716A_{21}x_{0}^{11}
			\\
			\hspace{2.5em}+646646A_{22}x_{0}^{12}-1144066A_{23}x_{0}^{13}+1961256A_{24}x_{0}^{14}-3268760A_{25}x_{0}^{15}+5311735A_{26}x_{0}^{16}
			\\
			\hspace{2.5em}-8436285A_{27}x_{0}^{17}+13123110A_{28}x_{0}^{18}
			\\
			B_{11}=A_{11}-12A_{12}x_0+78A_{13}x_{0}^{2}-364A_{14}x_{0}^{3}+1365A_{15}x_{0}^{4}-4368A_{16}x_{0}^{5}+12376A_{17}x_{0}^{6}
			\\
			\hspace{2.5em}-31824A_{18}x_{0}^{7}+75582A_{19}x_{0}^{8}-167960A_{20}x_{0}^{9}+352716A_{21}x_{0}^{10}
			\\
			\hspace{2.5em}-705432A_{22}x_{0}^{11}+1352078A_{23}x_{0}^{12}-2496144A_{24}x_{0}^{13}
			\\
			\hspace{2.5em}+4457400A_{25}x_{0}^{14}-7726160A_{26}x_{0}^{15}+13037895A_{27}x_{0}^{16}-21474180A_{28}x_{0}^{17}
		\end{array} \right.
	\end{aligned}
\end{align*}

\setcounter{equation}{6}
\begin{align}
	\begin{aligned}
		\left\{ \begin{array}{l}			
			B_{12}=A_{12}-13A_{13}x_0+91A_{14}x_{0}^{2}-455A_{15}x_{0}^{3}+1820A_{16}x_{0}^{4}-6188A_{17}x_{0}^{5}+18564A_{18}x_{0}^{6}
			\\
			\hspace{2.5em}-50388A_{19}x_{0}^{7}+125970A_{20}x_{0}^{8}-293930A_{21}x_{0}^{9}+646646A_{22}x_{0}^{10}
			\\
			\hspace{2.5em}-1352078A_{23}x_{0}^{11}+2704156A_{24}x_{0}^{12}-5200300A_{25}x_{0}^{13}
			\\
			\hspace{2.5em}+9657700A_{26}x_{0}^{14}-17383860A_{27}x_{0}^{15}+30421755A_{28}x_{0}^{16}			
			\\
			B_{13}=A_{13}-14A_{14}x_0+105A_{15}x_{0}^{2}-560A_{16}x_{0}^{3}+2380A_{17}x_{0}^{4}-8568A_{18}x_{0}^{5}+27132A_{19}x_{0}^{6}
			\\
			\hspace{2.5em}-77520A_{20}x_{0}^{7}+203490A_{21}x_{0}^{8}-497420A_{22}x_{0}^{9}+1144066A_{23}x_{0}^{10}
			\\
			\hspace{2.5em}-2496144A_{24}x_{0}^{11}+5200300A_{25}x_{0}^{12}-10400600A_{26}x_{0}^{13}
			\\
			\hspace{2.5em}+20058300A_{27}x_{0}^{14}-37442160A_{28}x_{0}^{15}
			\\
			B_{14}=A_{14}-15A_{15}x_0+120A_{16}x_{0}^{2}-680A_{17}x_{0}^{3}+3060A_{18}x_{0}^{4}-11628A_{19}x_{0}^{5}+38760A_{20}x_{0}^{6}
			\\
			\hspace{2.5em}-116280A_{21}x_{0}^{7}+319770A_{22}x_{0}^{8}-817190A_{23}x_{0}^{9}+1961256A_{24}x_{0}^{10}-4457400A_{25}x_{0}^{11}
			\\
			\hspace{2.5em}+9657700A_{26}x_{0}^{12}-20058300A_{27}x_{0}^{13}+40116600A_{28}x_{0}^{14}
			\\
			B_{15}=A_{15}-16A_{16}x_0+136A_{17}x_{0}^{2}-816A_{18}x_{0}^{3}+3876A_{19}x_{0}^{4}-15504A_{20}x_{0}^{5}+54264A_{21}x_{0}^{6}
			\\
			\hspace{2.5em}-170544A_{22}x_{0}^{7}+490314A_{23}x_{0}^{8}-1307504A_{24}x_{0}^{9}+3268760A_{25}x_{0}^{10}-7726160A_{26}x_{0}^{11}
			\\
			\hspace{2.5em}+17383860A_{27}x_{0}^{12}-37442160A_{28}x_{0}^{13}
				\end{array} \right.
		\end{aligned}
	\end{align}

\setcounter{equation}{6}

		\begin{align*}
			\begin{aligned}
			\left\{ \begin{array}{l}
			B_{16}=A_{16}-17A_{17}x_0+153A_{18}x_{0}^{2}-969A_{19}x_{0}^{3}+4845A_{20}x_{0}^{4}-20349A_{21}x_{0}^{5}+74613A_{22}x_{0}^{6}
			\\
			\hspace{2.5em}-245157A_{23}x_{0}^{7}+735471A_{24}x_{0}^{8}-2042975A_{25}x_{0}^{9}+5311735A_{26}x_{0}^{10}
			\\
			\hspace{2.5em}-13037895A_{27}x_{0}^{11}+30421755A_{28}x_{0}^{12}
			\\
			B_{17}=A_{17}-18A_{18}x_0+171A_{19}x_{0}^{2}-1140A_{20}x_{0}^{3}+5985A_{21}x_{0}^{4}-26334A_{22}x_{0}^{5}+100947A_{23}x_{0}^{6}
			\\
			\hspace{2.5em}-346104A_{24}x_{0}^{7}+1081575A_{25}x_{0}^{8}-3124550A_{26}x_{0}^{9}+8436285A_{27}x_{0}^{10}-21474180A_{28}x_{0}^{11}			
			\\
			B_{18}=A_{18}-19A_{19}x_0+190A_{20}x_{0}^{2}-1330A_{21}x_{0}^{3}+7315A_{22}x_{0}^{4}-33649A_{23}x_{0}^{5}+134596A_{24}x_{0}^{6}
			\\
			\hspace{2.5em}-480700A_{25}x_{0}^{7}+1562275A_{26}x_{0}^{8}-4686825A_{27}x_{0}^{9}+13123110A_{28}x_{0}^{10}
			\\
			B_{19}=A_{19}-20A_{20}x_0+210A_{21}x_{0}^{2}-1540A_{22}x_{0}^{3}+8855A_{23}x_{0}^{4}-42504A_{24}x_{0}^{5}+177100A_{25}x_{0}^{6}
			\\
			\hspace{2.5em}-657800A_{26}x_{0}^{7}+2220075A_{27}x_{0}^{8}-6906900A_{28}x_{0}^{9}					
		\end{array} \right.
		\end{aligned}
	\end{align*}
			
\setcounter{equation}{6}

\begin{align*}
	\begin{aligned}
	\left\{ \begin{array}{l}			
			B_{20}=A_{20}-21A_{21}x_0+231A_{21}x_{0}^{2}-1771A_{23}x_{0}^{3}+10626A_{24}x_{0}^{4}-53130A_{25}x_{0}^{5}+230230A_{26}x_{0}^{6}
			\\
			\hspace{2.5em}-888030A_{27}x_{0}^{7}+3108105A_{28}x_{0}^{8}\\
			B_{21}=A_{21}-22A_{22}x_0+253A_{23}x_{0}^{2}-2024A_{24}x_{0}^{3}+12650A_{25}x_{0}^{4}-65780A_{26}x_{0}^{5}+296010A_{27}x_{0}^{6}
			\\
			\hspace{2.5em}-1184040A_{28}x_{0}^{7}
			\\
			B_{22}=A_{22}-23A_{23}x_0+276A_{24}x_{0}^{2}-2300A_{25}x_{0}^{3}+14950A_{26}x_{0}^{4}-80730A_{27}x_{0}^{5}+376740A_{28}x_{0}^{6}
			\\
			B_{23}=A_{23}-24A_{24}x_0+300A_{25}x_{0}^{2}-2600A_{26}x_{0}^{3}+17550A_{27}x_{0}^{4}-98280A_{28}x_{0}^{5}
		\end{array} \right.
     \end{aligned}
\end{align*}

\setcounter{equation}{6}
\begin{align*}
	\begin{aligned}
		\left\{ \begin{array}{l}			
			B_{24}=A_{24}4-25A_{25}x_0+325A_{26}x_{0}^{2}-2925A_{27}x_{0}^{3}+20475A_{28}x_{0}^{4}\\
			B_{25}=A_{25}-26A_{26}x_0+351A_{27}x_{0}^{2}-3276A_{28}x_{0}^{3}
			\\
			B_{26}=A_{26}-27A_{27}x_0+378A_{28}x_{0}^{2}
			\\
			B_{27}=A_{27}-28A_{28}x_0
			\\
			B_{28}=A_{28}					
		\end{array} \right.
	\end{aligned}
\end{align*}

\begin{figure}[htb]
	\includegraphics[scale=0.6]{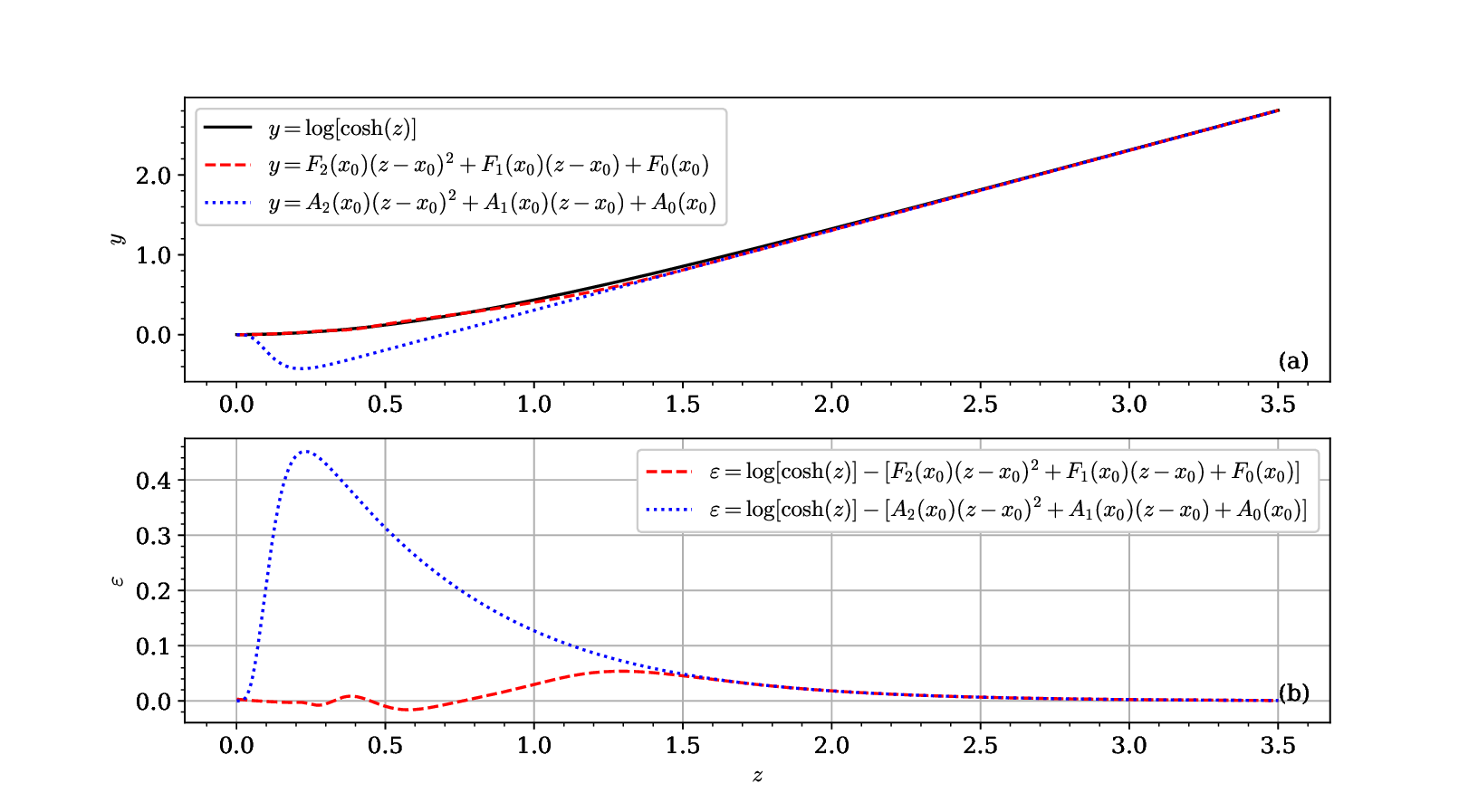}
	\caption{\footnotesize (a) Comparison of $\log[\cosh (z)]$ (solid line) with a 2-order series of $F_2(x_0)(z-x_0)^2+F_1(x_0)(z-x_0)+F_0 (x_0)$ (red dashed line), together with a 2-order Taylor series of $A_2(x_0)(z-x_0)^2+A_1(x_0)(z-x_0)+A_0 (x_0)$ (blue dotted line); (b) Variation of absolute error $\varepsilon$ with $z$. $x_0=10\sqrt{2}z$; $F_0(x_0)$, $F_1(x_0)$, and $F_2(x_0)$ are from (\ref{SCeq4_2})-(\ref{SCeq4_5}).}
	\label{fig1}
\end{figure}

Fig. \ref{fig1} shows the approximation to the function $\log[\cosh (z)]$ with a 2-order series of $F_2(z-x_0)^2+F_1(z-x_0)+F_0$ ($x_0=\sqrt{2}\eta z=10\sqrt{2}z$; Here $\eta =10$ for convenience) in which a 28-order Taylor series is used in (\ref{SCeq4_2})-(\ref{SCeq4_5}), and with a 2-order Taylor series of $A_2(x_0)(z-x_0)^2+A_1(x_0)(z-x_0)+A_0 (x_0)$, respectively. It can be found that: (1) $F_2(z-x_0)^2+F_1(z-x_0)+F_0$ can approximate well $\log[\cosh (z)]$ when $z\in [0,3.5]$, while  $A_2(x_0)(z-x_0)^2+A_1(x_0)(z-x_0)+A_0 (x_0)$ can represent $\log[\cosh (z)]$ only when $z$ is very small or large; (2) For $z\in [0, 3.5]$, the absolute errors $|\varepsilon_{_F}|$ from $F_2(z-x_0)^2+F_1(z-x_0)+F_0$ are less than those $|\varepsilon_{_A}|$ from $A_2(z-x_0)^2+A_1(z-x_0)+A_0$, except when $z$ is very small; (3) $|\varepsilon_{_F}|_{_{\rm max}} < 0.06$ when $z\in [0, 3.5]$, while $|\varepsilon_{_A}|_{_{\rm max}} > 0.4$; (4) $|\varepsilon_{_F}|_{_{\rm max}} < 0.009$ when $z\in [0, 0.5]$. Therefore, $F_2(z-x_0)^2+F_1(z-x_0)+F_0$ in Fig. \ref{fig1} can represent $\log[\cosh (z)]$ with $|\varepsilon_{_F}|_{_{\rm max}} < 0.06$ when $z\in [0, \infty]$, and with $|\varepsilon_{_F}|_{_{\rm max}} < 0.009$ when $z\in [0, 0.5]$. The latter are within the acceptable range and can be used for calculating $z_c$.

And now,

\setcounter{equation}{7}

\begin{align}\label{SCeq5}
	\begin{aligned}
		f\left( \lambda _p,t \right) &=-\frac{1}{2}\lambda _pt^2+\frac{2}{\pi}\int_0^{\pi /2}{\log \left( \cosh \left( \sqrt{2}\lambda _pt\sin \xi \right) \right) \mathrm{d}\xi}\\
		&=f\left( \bar{\lambda}_p,t^{\prime};z,\eta \right)\\
		&=\left( -\frac{1}{2}z\eta \right) \bar{\lambda}_pt^{\prime2}+\frac{2}{\pi}\int_0^{\pi /2}{\log \left\{ \cosh \left[ \left( \sqrt{2}z\eta \right) \bar{\lambda}_pt^{\prime}\sin \xi \right] \right\} \mathrm{d}\xi}\\
		&\approx \left( -\frac{1}{2}z\eta\,\bar{\lambda}_p+\frac{1}{2}F_2x_{0}^{2}\bar{\lambda}_{p}^{2} \right) t^{\prime2}-\frac{2}{\pi}\left( 2F_2x_{0}^{2}-F_1x_0 \right) \bar{\lambda}_pt^{\prime}+F_2x_{0}^{2}-F_1x_0+F_0&\hspace{1.5em}\\
	\end{aligned}
\end{align}

\begin{equation}\label{SCeq6}
	\frac{\mathrm{d}f\left( \bar{\lambda}_p,t^{\prime};z,\eta \right)}{\mathrm{d}t^{\prime}}=\left( -z\eta\bar{\lambda}_p+F_2x_{0}^{2}\bar{\lambda}_{p}^{2} \right) t^{\prime}-\frac{2}{\pi}\left( 2F_2x_{0}^{2}-F_1x_0 \right) \bar{\lambda}_p
\end{equation}

From that ${\mathrm{d}f\left( \bar{\lambda }_p,t^\prime;z,\eta \right)}/{\mathrm{d}t^\prime}=0$ we can get,

\begin{equation}\label{SCeq7}
	t_{sp}^{\prime}=\frac{\frac{2}{\pi}\left( F_1x_0-2F_2x_{0}^{2} \right)}{z\eta-F_2x_{0}^{2}\bar{\lambda}_p}
\end{equation}

Thus,

\begin{align}\label{SCeq8}
	\begin{aligned}
		f\left( \lambda _p,t_{sp} \right) &=f\left( \bar{\lambda}_p,t_{sp}^{\prime};z,\eta \right)\\
		&= \frac{\frac{2}{\pi ^2}\left( F_1x_0-2F_2x_{0}^{2} \right) ^2\bar{\lambda}_p}{z\eta -F_2x_{0}^{2}\bar{\lambda}_p}+F_2x_{0}^{2}-F_1x_0+F_0\hspace{1.5em}\\
		&\approx \frac{\frac{2}{\pi ^2}\left( F_1x_0-2F_2x_{0}^{2} \right) ^2\bar{\lambda}_p}{z\eta} +F_2x_{0}^{2}-F_1x_0+F_0
	\end{aligned}	
\end{align}

And, 

\begin{align}\label{SCeq9}
	\begin{aligned}
		\sum_{p=2}^N{f\left( \lambda _p,t_{sp} \right)}&=\sum_{p=2}^N{f\left( \bar{\lambda }_p,t_{sp}^{\prime}; z,\eta \right)}
		\\
		&\approx \sum_{p=2}^N \frac{\frac{2}{\pi ^2}\left( F_1x_0-2F_2x_{0}^{2} \right) ^2\bar{\lambda}_p}{z\eta }+F_2x_{0}^{2}-F_1x_0+F_0 
    \end{aligned}
\end{align}

And,

\begin{equation}\label{SCeq11}
\frac{\mathrm{d}^2f\left( \bar{\lambda}_p,t^{\prime};z,\eta  \right)}{\mathrm{d}t^{\prime2}}=z\eta \bar{\lambda}_p\left( -1+4zF_2\frac{\eta\bar{\lambda}_p}{2} \right) =z\lambda _{p}^{\prime}\left( -1+4zF_2\frac{\lambda _{p}^{\prime}}{2} \right) 
\end{equation} 

\section{Calculating $z_c$ for 2D and 3D Ising model by graphing}\label{ApendDD}

\begin{figure}[htb]
	\includegraphics[scale=0.8]{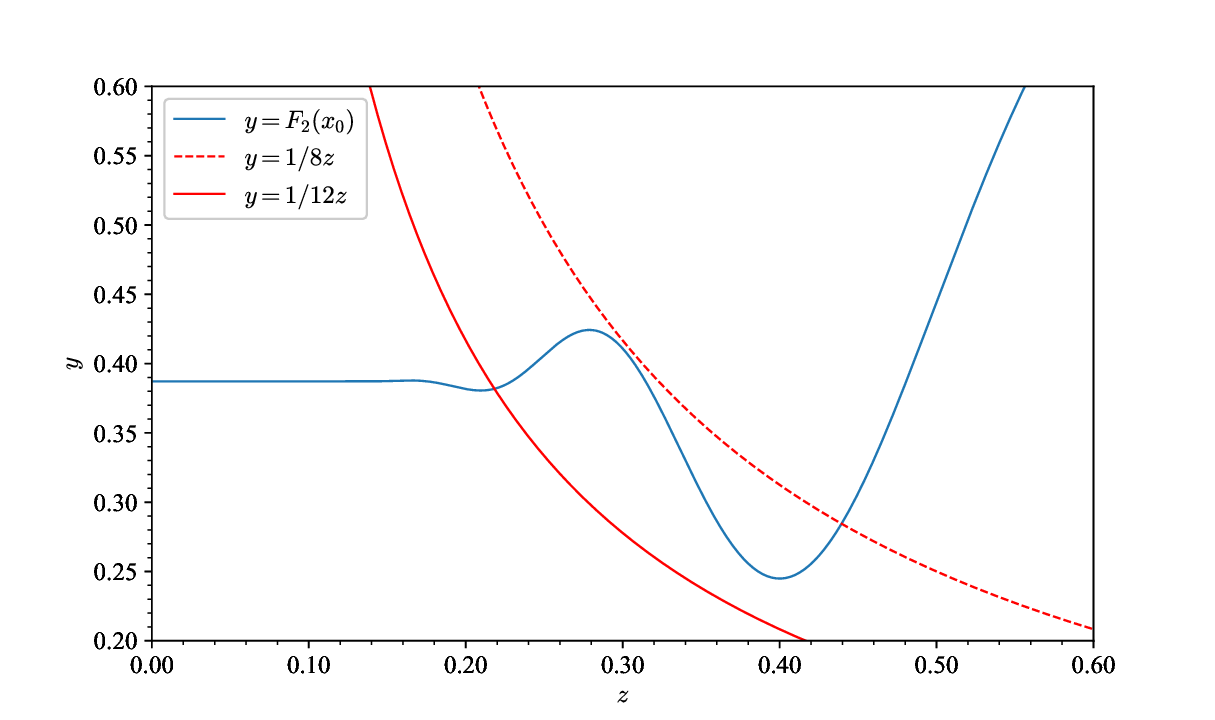}
	\caption{\footnotesize Variations of three curves ( $y=F_2(x_0)=F_2(10\sqrt{2}z)$, $y=1/8z$ and $y=1/12z$) with $z$. $F_2(10\sqrt{2}z)$ is calculated step by step using (\ref{SCeq4_2})-(\ref{SCeq4_5}). From these curves it can be found that $z_c\approx 0.439$ for 2D Ising model and that $z_c \approx 0.218$ for 3D Ising model, respectively.}
	\label{fig2}
\end{figure}

With (\ref{SCeq16_2d_new}) and (\ref{SCeq16_new}), we can estimate the critical inverse temperatures $z_c$ for 2D and 3D Ising model, respectively. For 2D model, $z_c$ is the intersection of two curves that $y=1/8z$ and that $y=F_2(x_0)=F_2(10\sqrt{2}z)$, while it is the intersection of two curves that $y=1/12z$ and that $y=F_2(10\sqrt{2}z)$ for 3D model. These three curves, when a 28-order Taylor series of $\log[\cosh(x)]$ is used in (\ref{SCeq4_2})-(\ref{SCeq4_5}), are shown in Fig. \ref{fig2}. From this figure we can obtain that $z_c\approx 0.439$ for 2D Ising model, and that $z_c\approx 0.218$ for 3D Ising model, respectively.

\setcounter{equation}{0} 
\section{$\sum_{i=2}^N{\cos\frac{2\pi}{N}(i-1)(p-1)}$ and $\sum_{i=2}^N{\sin\frac{2\pi}{N}(i-1)(p-1)}$}\label{ApendF}

The identity (\ref{eq_app_1}) and (\ref{h_eq_app_1}) will be used in the following,

\begin{equation}\label{h_eq_app_1}
	\sum_{k=1}^N{\sin kx}=\frac{\sin\frac{N+1}{2}x \sin\frac{N}{2}x }{\sin\frac{x}{2}}
\end{equation}

\begin{align}\label{h_eq_app_2}
	\begin{aligned}
		&\sum_{i=2}^N{\sin \frac{2\pi}{N}\left( p-1 \right) \left( i-1 \right)}\\
		&\,\,=\sum_{k=1}^{N-1}{\sin \left[ \frac{2\pi}{N}\left( p-1 \right) \right] k}\\
		&\,\,=-\sin N\left[ \frac{2\pi}{N}\left( p-1 \right) \right] +\sum_{k=1}^N{\sin \left[ \frac{2\pi}{N}\left( p-1 \right) \right] k}\\
%
%
		&\,\,=\frac{\sin \frac{N+1}{2}\left[ \frac{2\pi}{N}\left( p-1 \right) \right] \sin \frac{N}{2}\left[ \frac{2\pi}{N}\left( p-1 \right) \right]}{\sin \frac{1}{2}\left[ \frac{2\pi}{N}\left( p-1 \right) \right]}\\
		\\
		&=\frac{\cos \left( p-1 \right) \pi \,\,\sin \frac{\left( p-1 \right) \pi}{N}\,\,\sin \left( p-1 \right) \pi}{\sin \frac{\left( p-1 \right) \pi}{N}}
		\\
		&=\cos \left( p-1 \right) \pi \,\,\sin \left( p-1 \right) \pi 
		\\
		& =0
	\end{aligned}
\end{align}

\begin{align}\label{h_eq_app_3}
	\begin{aligned}
		&\sum_{i=2}^N{\cos \frac{2\pi}{N}\left( p-1 \right)}\left( i-1 \right) 
		\\
		&\,\,=\sum_{k=1}^{N-1}{\cos \left[ \frac{2\pi}{N}\left( p-1 \right) \right] k}
		\\
		&\,\,=\left\{ \sum_{k=0}^N{\cos \left[ \frac{2\pi}{N}\left( p-1 \right) \right] k} \right\} -\cos \left[ \frac{2\pi}{N}\left( p-1 \right) \right] 0-\cos \left[ \frac{2\pi}{N}\left( p-1 \right) \right] N
		\\
		&\,\,=\frac{1}{2}\left\{ 1+\frac{\sin \left( N+\frac{1}{2} \right) \left[ \frac{2\pi}{N}\left( p-1 \right) \right]}{\sin \left[ \frac{\pi}{N}\left( p-1 \right) \right]} \right\} -2
		\\
		&\,\,=-1
	\end{aligned}
\end{align}

%
%
%
%

\setcounter{equation}{0}
\section{$\mathcal{Z}$ and $I_p$ for Ising model in a non-zero external magnetic field}\label{ApendH}

Using the following $N$-dimensional Gaussian integrals for a positive definite matrix $A$ (namely Hubbard-Stratonovich transformation in section \ref{HS_0h}),

\begin{align}\label{eq29}
	\begin{aligned}
		\int_{-\infty}^{+\infty}\int_{-\infty}^{+\infty}{\cdots}\int_{-\infty}^{+\infty}{\prod_{i=1}^N{\frac{\mathrm{d}x_i}{\sqrt{2\pi}}}}\exp \left( -\frac{1}{2}x_iA_{ij}x_j+x_is_i \right)
		=\left[ \det  \mathbf{A} \right] ^{-1/2}\exp \left( \frac{1}{2}s_i{A_{ij}^{-1}}s_j \right)
	\end{aligned}
\end{align}

and writing the Ising model (\ref{eq28}) in the following form (Let $K_{ij}={A_{ij}^{-1}}$), we have,

\begin{align}\label{eq30}
	\begin{aligned}
		\exp \left( \frac{1}{2}s_iK_{ij}s_j \right) 
		&=\frac{\int_{-\infty}^{+\infty}\int_{-\infty}^{+\infty}{\cdots}\int_{-\infty}^{+\infty}{\prod_{i=1}^N{\frac{\mathrm{d}x_i}{\sqrt{2\pi}}}}\exp \left( -\frac{1}{2}x_iK_{ij}^{-1}x_j+x_is_i \right)}{\int_{-\infty}^{+\infty}\int_{-\infty}^{+\infty}{\cdots}\int_{-\infty}^{+\infty}{\prod_{i=1}^N{\frac{\mathrm{d}x_i}{\sqrt{2\pi}}}}\exp \left( -\frac{1}{2}x_iK_{ij}^{-1}x_j \right)}
	\end{aligned}
\end{align}

Then the partition function is,

\begin{align}\label{eq31}
	\begin{aligned}
		\mathcal{Z} &=\frac{\int_{-\infty}^{+\infty}\int_{-\infty}^{+\infty}{\cdots}\int_{-\infty}^{+\infty}{\prod_{i=1}^N{\frac{\mathrm{d}x_i}{\sqrt{2\pi}}}}\exp \left( -\frac{1}{2}x_iK_{ij}^{-1}x_j \right) \sum_{\left\{ s_i \right\}}{\exp \left[ s_i\left( x_i+\tilde{h}_i \right) \right]}}{\int_{-\infty}^{+\infty}\int_{-\infty}^{+\infty}{\cdots}\int_{-\infty}^{+\infty}{\prod_{i=1}^N{\frac{\mathrm{d}x_i}{\sqrt{2\pi}}}}\exp \left( -\frac{1}{2}x_iK_{ij}^{-1}x_j \right)}
		\\
		&\xlongequal[x_i=\grave{x}_i]{\grave{x}_i=x_i+\tilde{h}_i}\frac{\int_{-\infty}^{+\infty}\int_{-\infty}^{+\infty}{\cdots}\int_{-\infty}^{+\infty}{\prod_{i=1}^N{\frac{\mathrm{d}x_i}{\sqrt{2\pi}}}}\exp \left[ -\frac{1}{2}\left( x_i-\tilde{h}_i \right) K_{ij}^{-1}\left( x_j-\tilde{h}_j \right) \right] \sum_{\left\{ s_i \right\}}{\exp \left( s_ix_i \right)}}{\int_{-\infty}^{+\infty}\int_{-\infty}^{+\infty}{\cdots}\int_{-\infty}^{+\infty}{\prod_{i=1}^N{\frac{\mathrm{d}x_i}{\sqrt{2\pi}}}}\exp \left[ -\frac{1}{2}\left( x_i-\tilde{h}_i \right) K_{ij}^{-1}\left( x_i-\tilde{h}_i \right) \right]}
		\\
		&\xlongequal{\phi _i={K_{ij}}^{-1}x_j}\frac{\int_{-\infty}^{+\infty}\int_{-\infty}^{+\infty}{\cdots}\int_{-\infty}^{+\infty}{\prod_{i=1}^N{\frac{\mathrm{d}\phi _i}{\sqrt{2\pi}}}}\exp \left[ -\frac{1}{2}\left( \phi _iK_{ij}\phi _j-2\tilde{h}_i\phi _i+h_iK_{ij}^{-1}\tilde{h}_j \right) \right] \sum_{\left\{ s_i \right\}}{\exp \left( s_iK_{ij}\phi _j \right)}}{\int_{-\infty}^{+\infty}\int_{-\infty}^{+\infty}{\cdots}\int_{-\infty}^{+\infty}{\prod_{i=1}^N{\frac{\mathrm{d}\phi _i}{\sqrt{2\pi}}}}\exp \left[ -\frac{1}{2}\left( \phi _iK_{ij}\phi _j-2\tilde{h}_i\phi _i+h_iK_{ij}^{-1}\tilde{h}_j \right) \right]}\\
	&=\frac{\int_{-\infty}^{+\infty}\int_{-\infty}^{+\infty}{\cdots}\int_{-\infty}^{+\infty}{\prod_{i=1}^N{\frac{\mathrm{d}\phi _i}{\sqrt{2\pi}}}}\exp \left( -\frac{1}{2}\phi _iK_{ij}\phi _j+\tilde{h}_i\phi _i \right) \sum_{\left\{ s_i \right\}}{\exp \left( s_iK_{ij}\phi _j \right)}}{\int_{-\infty}^{+\infty}\int_{-\infty}^{+\infty}{\cdots}\int_{-\infty}^{+\infty}{\prod_{i=1}^N{\frac{\mathrm{d}\phi _i}{\sqrt{2\pi}}}}\exp \left( -\frac{1}{2}\phi _iK_{ij}\phi _j+\tilde{h}_i\phi _i \right)}
	\\
	& \\
	&=\left[ \frac{\det  \mathbf{K}}{\left( 2\pi \right) ^N} \right] ^{1/2}\exp \left( -\frac{1}{2}\tilde{h}_iK_{ij}^{-1}\tilde{h}_j \right) \int_{-\infty}^{+\infty}{\cdots}\int_{-\infty}^{+\infty}{\prod_{i=1}^N{\mathrm{d}\phi _i}}\exp \left( -\frac{1}{2}\phi _iK_{ij}\phi _j+\tilde{h}_i\phi _i \right) \\
	& \hspace{22em}\times \sum_{\left\{ s_i \right\}}{\exp \left( s_iK_{ij}\phi _j \right)}	
\end{aligned}
\end{align}
where $\frac{\grave{x}_i=x_i+\tilde{h}_i}{x_i=\grave{x}_i}$ means that firstly $\grave{x}_i=x_i+\tilde{h}_i$, and then $x_i=\grave{x}_i$ without causing confusion. $\sum_{\left\{ s_i \right\}}=\sum_{\left\{ s_i=\pm 1 \right\}}$. 

According to (\ref{eq5}), we know the item, $$\sum_{\left\{ s_i \right\}}{\exp \left( s_iK_{ij}\phi _j \right)}$$ in (\ref{eq31}) can be evaluated as the follows, 

\begin{align}\label{eq32}
\begin{aligned}
	\sum_{\left\{ s_i \right\}}{\exp \left({s_iK_{ij}\phi _j} \right)}
	&=\sum_{\left\{ s_i=\pm 1 \right\}}{\exp \left( \sum_p{s_iV_{ip}V_{pi}K_{ij}V_{jp}V_{pj}\phi _j} \right)}\\
	&=\prod_p{\sum_{\left\{ s_i=\pm 1 \right\}}{\prod_i{\exp \left[ s_i\left( V_{ip}\lambda _py_p \right) \right]}}}\\
	&=\prod_p{\exp \left\{ \sum_i{\log \left[\cosh \left( V_{ip}\lambda _py_p \right) \right]} \right\}}
\end{aligned}
\end{align} 

With $$\tilde{h}_i\phi _i=\tilde{h}_iV_{ip}V_{pi}\phi _i=\tilde{h}_iV_{ip}y_p$$ and (\ref{eq32}), we have, 

\begin{align}\label{eq33}
\begin{aligned} 
	\mathcal{Z}&\propto \int_{-\infty}^{+\infty}\int_{-\infty}^{+\infty}\dotsi\int_{-\infty}^{+\infty}\prod_{p=1}^N \exp \left( -\frac{1}{2}\lambda _py_{p}^{2}+\sum_i{\tilde{h}_iV_{ip}y_p} \right)\\
	&\hspace{7em}\times  \exp \left\{ \sum_i{\log \left[\cosh \left( V_{ip}\lambda _py_p \right) \right]} \right\} \mathrm{d}y_p
	\\
	&=\prod_{p=1} \int_{-\infty}^{+\infty}\exp \left( -\frac{1}{2}\lambda _py_{p}^{2}+\sum_i{\tilde{h}_iV_{ip}y_p} \right) \\
	&\hspace{7em}\times \left\{ \exp  \sum_i{\log \left[\cosh \left( V_{ip}\lambda _py_p \right) \right]} \right\}\mathrm{d}y_p
	\\
	&=\prod_{p=1}{I_p}		
\end{aligned}
\end{align}

And $I_p$ now is,

\begin{align}\label{eq34}
\begin{aligned} 
	I_p&=\int_{-\infty}^{+\infty}{\exp \left( -\frac{1}{2}\lambda _py_{p}^{2}+\sum_i{\tilde{h}_iV_{ip}y_p} \right) \exp \left\{ \sum_i{\log \left[ \cosh \left( V_{ip}\lambda _py_p \right) \right]} \right\}  \mathrm{d}y_p}\\
	&\xlongequal{y_p=yN^{-1/2}}\int_{-\infty}^{+\infty}\exp \left( -\frac{1}{2}\lambda _p\frac{y^2}{N} \right) \exp \left( \sum_i{\tilde{h}_i\frac{\overline{V}_{ip}y}{N}} \right) \\
	&\hspace{10em} \times \exp \left\{ \sum_i{\log \left[ \cosh \left( \frac{\overline{V}_{ip}\lambda _py}{N} \right) \right]} \right\} \mathrm{d}\frac{y}{N^{1/2}}\\
	&\xlongequal{y=tN} N^{1/2}\int_{-\infty}^{+\infty}{\exp \left( -\frac{1}{2}\lambda _pNt^2 \right) }
	\exp \left\{ \sum_{i}{\tilde{h}_i\left[ \cos \frac{2\pi}{N}\left( i-1 \right) \left( p-1 \right) +\sin \frac{2\pi}{N}\left( i-1 \right) \left( p-1 \right) \right] t} \right\} \,\,        \\
	&\hspace{2em} \,\, \times \exp \left\{ \sum_{i}{\log \left\{ \cosh \left\{ \left[ \cos \frac{2\pi}{N}\left( i-1 \right) \left( p-1 \right) +\sin \frac{2\pi}{N}\left( i-1 \right) \left( p-1 \right) \right] \lambda _pt \right\} \right\}} \right\}  \mathrm{d}t\\
	& \\
	& =N^{1/2}\int_{-\infty}^{+\infty}{\exp \left( -\frac{1}{2}\lambda _pNt^2 \right)}\exp \left( Q_1 \right) \,\,\exp \left( Q_2 \right) 
\end{aligned}
\end{align}

When $p=1$, 

\begin{align}
\begin{aligned}
	Q_1&=\sum_{i}{\tilde{h}_i\left[ \cos \frac{2\pi}{N}\left( i-1 \right) \left( p-1 \right) +\sin \frac{2\pi}{N}\left( i-1 \right) \left( p-1 \right) \right] t}\\
	&=N\tilde{h}t \ \  (\mbox{ie.} \ N\tilde{h}_1t)
\end{aligned}
\end{align}

\begin{align}\label{Q2}
\begin{aligned}
	Q_2&=\sum_{i}{\log \left\{ \cosh \left\{ \left[ \cos \frac{2\pi}{N}\left( i-1 \right) \left( p-1 \right) +\sin \frac{2\pi}{N}\left( i-1 \right) \left( p-1 \right) \right] \lambda _pt \right\} \right\}}\\
	&=N\log[\cosh ( \lambda _1t )]
\end{aligned}
\end{align}

\begin{align}\label{eq34_add}
\begin{aligned}
	I_1& =N^{1/2}\int_{-\infty}^{+\infty}\exp\left\{ N\left[ -\frac{1}{2}\lambda _1 t^2 + \tilde{h}_1 t+\log [\cosh (\lambda _1 t)]\right]\right\}{\rm d}t
\end{aligned}
\end{align}

When $p>1$, $Q_1$ can be obtained according to (\ref{h_eq_app_2}) and (\ref{h_eq_app_3}),

\begin{align}\label{eq35}
\begin{aligned}
	Q_1&=\sum_{i}{\tilde{h}_i\left[ \cos \frac{2\pi}{N}\left( i-1 \right) \left( p-1 \right) +\sin \frac{2\pi}{N}\left( i-1 \right) \left( p-1 \right) \right] t}
	\\
	&={\tilde{h}_1t + \sum_{i=2}{\tilde{h}_i\left[ \cos \frac{2\pi}{N}\left( i-1 \right) \left( p-1 \right) +\sin \frac{2\pi}{N}\left( i-1 \right) \left( p-1 \right) \right] t}}\\
	&=0
\end{aligned}
\end{align}

Since $Q_2$ is the same to (\ref{eq13}),

\begin{align}\label{eq36}
\begin{aligned}
	I_p&=N^{1/2}\int_{-\infty}^{+\infty}\exp \left[N\left( -\frac{1}{2}\lambda _pt^2+\mathcal{I}(\lambda_p,t)\right) \right] \mathrm{d}t	
\end{aligned}
\end{align}

So the partition function now is,

\begin{align}\label{eq37}
\begin{aligned}
	\mathcal{Z} &=\left[ \frac{\det  \mathbf{K}}{\left( 2\pi \right) ^N} \right] ^{1/2}\exp \left( -\frac{1}{2}\tilde{h}_iK_{ij}^{-1}\tilde{h}_j \right) I_1 \prod_{p=2}{I_p}
	\\
	&=\left[ \frac{\det  \mathbf{K}}{\left( 2\pi \right) ^N} \right] ^{1/2}\exp \left( -\frac{1}{2}\sum_p{\frac{\tilde{h}^2}{\lambda _p}} \right) I_1 \\
	&\hspace{4em}\times\prod_{p=2}{ N^{1/2}\int_{-\infty}^{+\infty} \exp \left[ N\left( -\frac{1}{2}\lambda _pt^2+\mathcal{I}(\lambda_p,t) \right) \right] \mathrm{d}t}		
\end{aligned}
\end{align}
where $\exp \left( -\frac{1}{2}\tilde{h}_iK_{ij}^{-1}\tilde{h}_j \right) 
=\exp \left( -\frac{1}{2}\sum_p{\frac{\tilde{h}^2}{\lambda _p}} \right)$. 

It can be found that the $\lambda_p$ in $\exp \left( -\frac{1}{2}\sum_p{{\tilde{h}^2}/{\lambda _p}} \right)$ above is in the denominator, which results in a non-regular $f(\lambda_p/2)$ (See details in Appendix \ref{ApendC}). To avoid this and according to (\ref{bar_K}), an extra parameter $\alpha$ should be added to the partition function (\ref{eq4}), ie.,

\begin{align}\label{eq4_add_alpha}
\begin{aligned}
	\mathcal{Z} &=\sum_{\left\{ s_i=\pm 1 \right\}}{\exp \left( \frac{1}{2}\sum_{ij}{K_{ij}s_is_j} \right) \exp \left( 2N\alpha \right)} \exp \left( -2N\alpha \right)
	\\
	&=\sum_{\left\{ s_i=\pm 1 \right\}}\exp \left( -2N\alpha \right) \left[ \frac{\det \mathbf{\widetilde{K}} }{\left( 2\pi \right) ^N} \right] ^{1/2}\int\limits_{-\infty}^{+\infty}\int\limits_{-\infty}^{+\infty}\dotsi\int\limits_{-\infty}^{+\infty}\prod_{k=1}^N\mathrm{d}\phi _k\\
	&\hspace{16em}\times\exp \left( -\frac{1}{2}\sum_{ij}{\phi _i\widetilde{K}_{ij}\phi _j+\sum_{ij}{s_i\widetilde{K}_{ij}\phi _j}} \right)		
\end{aligned}
\end{align}

Finally, $I_1$ and $I_p$ ($p>1$) now are,

\begin{align}\label{eq_non0h_add1}
\begin{aligned}
	I_1&=N^{1/2}\int_{-\infty}^{+\infty}{\exp}\left\{ N\left[ -\frac{1}{2}\lambda _1t^2+\tilde{h}t+\log\cosh [(\lambda _1t)] \right] \right\} \mathrm{d}t
	\\
	&\sim 2\exp \left\{ N\left[ -\frac{1}{2}\lambda _1t_{s1}^{2}+\tilde{h}t_{s1}+\log \left[\cosh \left( \lambda _1t_{s1} \right) \right] \right] \right\} \sqrt{\frac{2\pi}{\left| \lambda _1-\lambda _{1}^{2}{\rm sech}^2\left( \lambda _1t_{s1} \right) \right|}}		
\end{aligned}
\end{align}
where $t_{s1}$ is now from that $t=\tanh \left( \lambda _1t \right) +\tilde{h}/\lambda _1$.

\begin{align}\label{eq_non0h_add2}
\begin{aligned}
	I_p&=N^{1/2}\int_{-\infty}^{+\infty} \exp \left[ N\left( -\frac{1}{2}\lambda _pt^2+\mathcal{I}(\lambda_p, t)\right) \right] \mathrm{d}t
	\\
	&\,\,   \sim \,\,2 \exp \left[ Nf\left( \lambda _p,t_{sp} \right) \right] \sqrt{\frac{2\pi}{\left| -\lambda _p+\mathcal{I}^{''}\left(\lambda _p,t \right) \right|}}		
\end{aligned}
\end{align}
where $t_{sp}$ is from that $-\lambda_pt+\mathcal{I}^{\prime}(\lambda_p, t) =0$. 

And the partition function $\mathcal{Z}$ is,

\begin{align}\label{eq_non0h_add3}
	\begin{aligned}
		\mathcal{Z}& =\exp \left( -2N\alpha \right) \left[ \frac{\det  \tilde{\mathbf{K}
		}}{\left( 2\pi \right) ^N} \right] ^{1/2}\exp \left( -\frac{1}{2}\sum_p{\frac{\tilde{h}^2}{\lambda _p}} \right) I_1\prod_{p=2}{I_p}
		\\
		&=\exp \left( -2N\alpha \right) \exp \left( -\frac{1}{2}\sum_p{\frac{\tilde{h}^2}{\lambda _p}} \right) 
		\\
		&\hspace{2em}  \times 2\exp \left\{ N\left[ -\frac{1}{2}\lambda _1t_{s1}^{2}+\tilde{h}t_{s1}+\log \left( \cosh \left( 2t_{s1}\frac{\lambda _1}{2} \right) \right) \right] \right\} \left| 1-\lambda _1{\rm sech}^2\left( \lambda _1t_{s1} \right) \right|^{-1/2}
		\\
		&\hspace{2em}  \times \prod_{p=2}{2 \exp[Nf(\lambda _p,t_{sp})]    \left| -1+\frac{\mathcal{I}^{''}\left(\lambda _p,t_{sp} \right)}{\lambda_p} \right|^{-1/2}}
	\end{aligned}
\end{align}

\vspace{5em}

\ \ 

\end{document}